\documentclass[12pt]{article}
\usepackage{amsmath}
\usepackage{amsthm}
\usepackage{amsfonts}
\usepackage[dvips]{epsfig}
\usepackage{graphicx}
\usepackage{amssymb}
\usepackage{cancel}
\usepackage{pstricks} 
\usepackage{lscape}
\usepackage{enumitem}
\usepackage{comment}
\usepackage{subcaption}

\usepackage[sort&compress,numbers, merge]{natbib}

\DeclareFontFamily{U}{mathx}{\hyphenchar\font45}
\DeclareFontShape{U}{mathx}{m}{n}{<-> mathx10}{}
\DeclareSymbolFont{mathx}{U}{mathx}{m}{n}

\newcommand{\beq}{\begin{equation}}
\newcommand{\eeq}{\end{equation}}
\newcommand{\bea}{\begin{eqnarray}}
\newcommand{\eea}{\end{eqnarray}}

\makeatletter
\newlength{\apb@width}
\newcommand{\autoparbox}[2][c]{\settowidth{\apb@width}{#2}\parbox[#1]{\apb@width}{#2}}

\newcommand{\Cen}[2]{%
  \ifmeasuring@
    #2%
  \else
    \makebox[\ifcase\expandafter #1\maxcolumn@widths\fi]{$\displaystyle#2$}%
  \fi
}
\makeatother

\definecolor{Orange}{cmyk}{0,0.61,0.87,0}
\definecolor{JungleGreen}{cmyk}{0.99,0,0.52,0}
\definecolor{OliveGreen}{cmyk}{0.64,0,0.95,0.40}
\definecolor{Brown}{cmyk}{0,0.81,1,0.60}
\definecolor{RoyalBlue}{cmyk}{0.71,0.53,0,0.12}

\allowdisplaybreaks[1]

\usepackage[colorlinks=true, linkcolor=OliveGreen, citecolor=RoyalBlue,
urlcolor=RoyalBlue]{hyperref}

\begin{document}

\def\a{\alpha}
\def\b{\beta}
\def\c{\varepsilon}
\def\d{\delta}
\def\e{\epsilon}
\def\f{\phi}
\def\g{\gamma}
\def\h{\theta}
\def\k{\kappa}
\def\l{\lambda}
\def\m{\mu}
\def\n{\nu}
\def\p{\psi}
\def\q{\partial}
\def\r{\rho}
\def\s{\sigma}
\def\t{\tau}
\def\u{\upsilon}
\def\v{\varphi}
\def\w{\omega}
\def\x{\xi}
\def\y{\eta}
\def\z{\zeta}
\def\D{\Delta}
\def\G{\Gamma}
\def\H{\Theta}
\def\L{\Lambda}
\def\F{\Phi}
\def\P{\Psi}
\def\S{\Sigma}

\def\o{\over}
\def\beq{\begin{eqnarray}}
\def\eeq{\end{eqnarray}}
\newcommand{\gsim}{ \mathop{}_{\textstyle \sim}^{\textstyle >} }
\newcommand{\lsim}{ \mathop{}_{\textstyle \sim}^{\textstyle <} }
\newcommand{\vev}[1]{ \left\langle {#1} \right\rangle }
\newcommand{\bra}[1]{ \langle {#1} | }
\newcommand{\ket}[1]{ | {#1} \rangle }
\newcommand{\EV}{ {\rm eV} }
\newcommand{\KEV}{ {\rm keV} }
\newcommand{\MEV}{ {\rm MeV} }
\newcommand{\GEV}{ {\rm GeV} }
\newcommand{\TEV}{ {\rm TeV} }
\def\diag{\mathop{\rm diag}\nolimits}
\def\Spin{\mathop{\rm Spin}}
\def\SO{\mathop{\rm SO}}
\def\O{\mathop{\rm O}}
\def\SU{\mathop{\rm SU}}
\def\U{\mathop{\rm U}}
\def\Sp{\mathop{\rm Sp}}
\def\SL{\mathop{\rm SL}}
\def\tr{\mathop{\rm tr}}

\def\IJMP{Int.~J.~Mod.~Phys. }
\def\MPL{Mod.~Phys.~Lett. }
\def\NP{Nucl.~Phys. }
\def\PL{Phys.~Lett. }
\def\PR{Phys.~Rev. }
\def\PRL{Phys.~Rev.~Lett. }
\def\PTP{Prog.~Theor.~Phys. }
\def\ZP{Z.~Phys. }

\vspace{0.5cm}
\begin{center}
{\bf \Large Flavor- and CP-safe explanation of $g_\mu-2$ anomaly}
\end{center}
\vspace{0.75cm}

\begin{center}
{\bf Jason~L.~Evans}$^{a}$,
{\bf Tsutomu~T.~ Yanagida}$^{a,b}$, and
{\bf Norimi Yokozaki}$^c$
\end{center}

\begin{center}
  {\em $^a$Tsung-Dao Lee Institute, Shanghai Jiao Tong University, Shanghai 200240, China}\\[0.2cm] 
  {\em $^b$Kavli IPMU (WPI), UTIAS, University of Tokyo, Kashiwa, Chiba 277-8583, Japan}\\[0.2cm] 
{\em $^c$Zhejiang Institute of Modern Physics and Department of Physics, Zhejiang University, Hangzhou, Zhejiang 310027, China}\\[0.2cm]

 \end{center}

\vspace{1cm}
\centerline{\bf ABSTRACT}
\vspace{0.2cm}

{\small 
Supersymmetry is still a viable explanation for the muon $g-2$ anomaly, if the sleptons and electroweak gauginos are $\mathcal{O}(100)$\,GeV. However, for supersymmetry breaking masses this light, the SUSY flavor and CP-problem are exacerbated. To address this issue, we consider a flavor-safe gauge mediated explanation of the muon $g-2$ with additional Higgs soft supersymmetry breaking mass parameters. The setup provides a generic parameter space within minimal gauge mediation. Furthermore, we show that the problematic CP violating phase can be dynamically suppressed. We find that gauge mediation models have large portions of parameter space where the muon $g-2$ can be explained at 1\,$\sigma$ level. The interplay between the slepton and the CP-odd Higgs masses also makes the majority of this model's parameter space testable at the LHC through searches for sleptons or additional Higgs bosons.
}

\vspace{0.5in}

\medskip
\noindent

\newpage

\section{Introduction}
The disagreement between the standard model prediction of $g_\mu-2$ and the measured value has a long history, with the discrepancy first showing up more than 20 years ago \cite{Muong-2:2001kxu}. The recent experimental measurement at Fermilab \cite{Muong-2:2021ojo}, when compared with recent theoretical predictions \cite{Aoyama:2020ynm,Colangelo:2018mtw, Hoferichter:2019mqg, Davier:2019can, Keshavarzi:2019abf}, has further increased the significance of this disagreement which is now at $4.2$ standard deviations.
Although not at the level of observation,\footnote{Some lattice calculations have weakened the significance of this deviation \cite{Borsanyi:2020mff} but are ongoing.} this deviation warrants our attention. 

Although explaining $g_\mu-2$ is not so difficult phenomenologically, it is non-trivial to find a model which is motivated. One of the best explanations for the deviation of $g_\mu-2$, which is motivated beyond just its prediction of $g_\mu-2$, is supersymmetry. In fact, $g_\mu-2$ merely constrains the mass parameters of supersymmetry, since it already contains the necessary particles and order one couplings with the muon to explain $g_\mu-2$~\cite{Lopez:1993vi, Chattopadhyay:1995ae, Moroi:1995yh}. Furthermore, these constraints on the spectra prefer light soft masses and so tend to support the other nice features of supersymmetry not contradict them.  For example, in \cite{Endo:2017zrj, Cox:2018qyi, Cox:2018vsv, Cox:2021nbo} it was shown that supersymmetric dark matter could be realized in models which explain $g_\mu-2$. And, of course, naturalness is unaffected by these constraints since it prefers light soft masses. Clearly, supersymmetry is a prime candidate for explaining $g_\mu-2$.  
The main challenge for explaining $g_\mu-2$ in supersymmetry is avoiding exclusion limits or theoretical problems. For example, the LHC slepton searches~\cite{ATLAS:2019lng,CMS:2020bfa}, the chargino/neutralino searches~\cite{ATLAS:2021moa, CMS:2021few, ATLAS:2021yqv} and the vacuum stability of the stau-Higgs potential~\cite{Hisano:2010re, Kitahara:2013lfa, Endo:2013lva} place strong constraints on the light needed slepton spectra. These constraints on the spectra are relatively easy to avoid if the correction to $g_\mu-2$ arises from a chargino loop.\footnote{
If the flavor universality of sleptons is violated and the staus are significantly heavier than the other sleptons at low energy, the LHC and vacuum stability constraints can be avoided even when $g_\mu-2$ anomaly is explained by a bino-loop~\cite{Ibe:2013oha,Endo:2013lva,Ibe:2019jbx}. A low scale stau mass, which is heavy enough and does not lead to flavor problems, can be realized through loop effects if the down-type Higgs boson soft mass is tachyonic and large at the ultraviolet scale~\cite{Yamaguchi:2016oqz, Yin:2016shg}.
}
An important theoretical consideration, which we do not consider here, is can $g_\mu-2$ be explained in models consistent with grand unification. It was shown that, at least in some unification models, it can be rectified \cite{Harigaya:2015kfa, Cox:2018vsv, Ellis:2021zmg,Ellis:2021vpp,Agashe:2022uih}.

Although recent studies of the supersymmetric explanation for $g_\mu-2$ are interesting, they tend to make assumptions about the sleptons and often neglect explaining key features of the needed spectrum like the lack of flavor and CP violation. These features of the spectrum are unavoidable when attempting to explaining $g_\mu-2$, since the slepton masses must be less than about 1\,TeV. If there are large amounts of flavor mixing or CP violation, the models are completely excluded. Since, generically, we expect order one flavor and CP violation in supersymmetry an explanation of their absence is needed. The flavor changing neutral current (FCNC) problem can be easily solved if supersymmetry breaking is mediated by gauge interactions.\footnote{
Gaugino mediation also offers a solution to the FCNC problem. However, it is in tension with the seesaw mechanism~\cite{seesaw_yanagida,Yanagida:1979gs,Gell-Mann:1979vob} if the sleptons are light enough to explain $g_\mu-2$~\cite{Nagai:2020xbq}.
}
Gauge mediation provides a reliable solution to the FCNC
problem based on four dimensional quantum field theory within perturbation theory.
In gauge mediated models, since sfermion masses are generated by gauge interactions, they have no generation mixing and viable models can be found \cite{Yanagida:2017dao,Bhattacharyya:2018inr} which avoid FCNC. CP violation, on the other hand, is not so simple of a problem to solve, even within gauge mediation since the gauginos and Higgs bilinear mass can still have order one CP violating phase.\footnote{
In \cite{Harigaya:2015kfa}, it was shown that both the flavor and CP-problem are solved within gaugino mediation. The CP-problem is solved by assuming a shift symmetry where the imaginary part of the SUSY breaking field receives a constant shift~\cite{Iwamoto:2014ywa}. See also ~\cite{Agashe:2022uih} for another construction of a CP-safe gaugino mediation model.  
}
In fact, the constraints on the CP violating phase associated with these parameter are so severe that their phases need to be less than about $10^{-5}$, as will be shown later, or the electric dipole moment of the electron will exceed experimental limits. Mindlessly setting the CP violation to zero in the SUSY breaking sector seems unnatural since we know that CP is almost maximally broken in the SM sector and generation of the CP violating phase in the Yukawa couplings from CP conserving models is non-trivial~\cite{Nelson:1983zb,Barr:1984qx,Dine:1993qm,Hiller:2002um,Evans:2020vil}.

An important question that remains for the supersymmetric explanation of $g_\mu-2$ is can an intelligent model produce the needed features in the slepton mass spectrum. This is the question we will examine here placing particular emphasis of the effects of CP violation. As we will take advantage of here, gauge mediation offers a dynamical means for suppressing the CP violating phases of the gaugino masses and Higgs $B$-term~\cite{Evans:2010ru}.\footnote{
In Ref.~\cite{Ibe:2021cvf}, another mechanism to suppress the CP violating phases is proposed within gauge mediation. 
} 
This solves the last major hurdle for explaining $g_\mu-2$, since gauge mediation already solves the FCNC problem and can be made consistent with all other relevant constraints. The aim of this work is to extend this original CP-safe gauge mediation model so that it can consistently explain the observed $g_\mu -2$ anomaly. In particular, we will show that the CP-safe mechanism proposed in Ref.~\cite{Evans:2010ru} becomes simpler and more generic by including anomaly mediated effects, which are usually neglected in gauge mediated models.


\section{$g-2$ anomaly and CP-violation}

In this section, we review the gauge mediation model in \cite{Bhattacharyya:2018inr} and its explanation of the $g_\mu-2$ anomaly. At the end of this section, we will show that this model, like most models, has problematically large CP violating phases unless the phases are taken to be unnaturally small.

The gauge mediation sector of the model in \cite{Bhattacharyya:2018inr} has an effective superpotential:
\begin{eqnarray}
W = (\kappa_L Z + M_{L})\Psi_L \Psi_{\bar L} 
+(\kappa_D Z + M_{D}) \Psi_D \Psi_{\bar D}, \label{eq:split_model}
\end{eqnarray}
where $Z$ is a SUSY breaking field whose $F$-component, $F_Z$, has a non-zero vacuum expectation value, and $\Psi_L$ and $\Psi_{\bar D}$ have the same SM charges as the left-handed leptons and down-type quarks, respectively, and $\Psi_{\bar L}$ and $\Psi_{D}$ are their conjugate fields. We have formulated the messenger sector so that $(\Psi_{\bar D}, \Psi_L)$ transform as $5^*$ of $SU(5)_{\rm GUT}$. However, we do not assume the existence of a grand unified theory (GUT). In fact, violation of the GUT relations, $\kappa_L/M_L \neq \kappa_D/M_D$,\footnote{
This combination is invariant under renormalization group evolution. Therefore, if we impose $k_L=k_D$ and $M_L=M_D$ at the ultraviolet scale (e.g., the GUT scale), $\kappa_L/M_L = \kappa_D/M_D$ holds at all scales. 
} 
is necessary or explaining $g_\mu-2$ and a 125 GeV Higgs boson mass becomes impossible due to competing requirements on the masses of sleptons and electroweak gauginos. This incongruence arises because the slepton masses are proportional to $\kappa_L \left<F_Z\right>/M_L$ while the squark and gluino masses are proportional to $\kappa_D \left<F_Z\right>/M_D$. The large stop mass needed to enhance the Higgs mass is inconsistent with light sleptons and the electroweak gaugino masses required for $g_\mu-2$ unless $|\kappa_D/M_D| \gg |\kappa_L/M_L|$. Therefore, the GUT relations on these couplings and masses must be broken such that $|\kappa_D/M_D| \gg |\kappa_L/M_L|$. (We have neglected the $U(1)_Y$ gauge coupling in this discussion since it is a subleading effect, see Eqs.~(\ref{eq:gaugino_masses}) and (\ref{eq:sfermion_masses}) below for the complete formulas.) Note, although one may consider messenger fields in other representations of $SU(5)_{\rm GUT}$ such as ${\bf 10} (+\overline{\bf 10})$ and a ${\bf 24}$, gauge mediated models with these messenger fields are not advantageous for explaining $g_\mu -2$~\cite{Yanagida:2017dao}.

Here, we show the necessity of additional contributions to the Higgs soft masses beyond those provided by gauge mediation. These additional soft masses are vital for decoupling the explanation of a 125 GeV Higgs boson mass from the value of the $\mu$ parameter. To explain a Higgs boson mass of 125 GeV in gauge mediation, which has a relatively small stop trilinear mass term, the stop mass needs to be at least 5\,TeV. Stop masses this large generate large radiative corrections to the Higgs soft masses. The electroweak symmetry breaking conditions then force $\mu\gtrsim$ 2\,TeV. With $\mu$ this large, the chargino-loop contribution to the $g_\mu-2$ is quite suppressed. The bino loop contribution, on the other hand, is enhanced by $\mu\tan\beta$ and so might seem like a viable alternative for explaining $g_\mu-2$. However, if the bino loop contribution to $g_\mu-2$ is to explain the $g_\mu-2$ anomaly, the sleptons need to be particularly light and $\mu \tan\beta$ needs to be rather large. 
Here, $\tan\beta$ is a ratio of the Higgs vacuum expectation values (VEVs).
This spectrum is, unfortunately, excluded by the wino mass limits ($\gsim660$\,GeV) of the LHC~\cite{ATLAS:2022rme} and/or the vacuum stability constraints from the stau-Higgs potential~\cite{Hisano:2010re, Kitahara:2013lfa, Endo:2013lva}. The strong limits on the wino mass arise because $|\kappa_D/M_D| \gg |\kappa_L/M_L|$ is needed to get the stop mass large enough to explain the Higgs boson mass. In this limit, the NLSP is an (almost) pure wino and so is long-lived. Long lived charge particles are strongly constrained by disappearing charged track searches at the LHC~\cite{CMS:2020atg,ATLAS:2022rme}. Furthermore, if the trilinear coupling between the staus and Higgs boson, which is proportional to $\mu \tan\beta$, is too large, a charge breaking minimum which is deeper than the electroweak symmetry breaking minimum emerges. For staus as light as a few hundred GeV, $\mu \tan\beta$ is strongly constrained by the requirement that the life-time of the electroweak symmetry breaking minimum is longer than the age of the universe. By taking $\mu$ of $\mathcal{O}(100)$\,GeV, both these problems are avoided. Then, the $g_\mu-2$ can be explained by the chargino loop contribution. To achieve a small $\mu$-term, our model generates an additional contribution to the soft SUSY breaking mass squared of the up-type Higgs, $m_{H_u}^2$, beyond the standard gauge mediated piece \cite{Ibe:2012qu,Yanagida:2017dao,Bhattacharyya:2018inr}. (See also Appendix \ref{sec:a1} for an example of how to generate $m_{H_u}^2$ and $m_{H_d}^2$.) 

As it turns out, the down type Higgs soft mass, $m_{H_d}^2$, also needs additional contributions to realize a viable model. In simple gauge mediation models, $m_{H_d}$ is equal to the left-handed slepton mass at the messenger scale, which needs to be of $\mathcal{O}(100)$\,GeV to explain $g_\mu-2$. This is not simultaneously consistent with electroweak symmetry breaking conditions or the LHC constraints. 

The problem with a small $m_{H_d}^2$ can be seen by examining one of the electroweak symmetry breaking conditions, which we write as 
\begin{eqnarray}
m_{H/A/H^\pm}^2 \simeq |B_{\mu}(Q)| (\tan\beta + \cot \beta) \simeq m_{H_d}^2(Q) + |\mu(Q)|^2,
\end{eqnarray}
where $H$, $A$ and $H^{\pm}$ are the CP-even heavy Higgs, CP-odd Higgs and charged Higgs, respectively. The renormalization scale $Q$ is taken to be around the stop mass scale, $m_{\rm stop}$; $B_\mu$ is defined by $V\ni B_\mu H_u H_d + h.c.$ with the superpotential $W=\mu H_u H_d$. Here, we neglect radiative corrections to the Higgs potential and terms suppressed by ${\mathcal O}(\tan^{-2}\beta)$. We have removed the argument of $B_\mu(Q)$ by the field redefinition of the Higgs fields, $|B_\mu(Q)|\exp(i \theta_{B_\mu}) \to |B_\mu(Q)|$, which rotates the phase of the $\mu$ term, $\mu \to \mu \exp(-i \theta_{B_\mu})$. In simple gauge mediation cases, $B_\mu(Q) = B(Q) \mu(Q)$ and $B(Q=m_{\rm stop})$ is radiatively generated through the gaugino masses. Since we need $\tan\beta\gtrsim O(10)$ to explain $g_\mu -2$, $m_{H/A}$ should be at least 1\,TeV to satisfy the LHC constraints~\cite{ATLAS:2020zms}. The larger $\tan\beta$ is the heavier $H$ and $A$ need to be. For example, $m_{H/A}$ should be $\gtrsim 2$\,TeV for $\tan\beta=60$. In this case, $m_{H_d}^2(Q=m_{\rm stop})$ needs to be $\mathcal{O}(10^6)$\,GeV$^2$. This is not achievable in gauge mediation models unless there is an additional source of $m_{H_d}^2$. Lastly, we also consider an additional source for the $B$-term, since a $B_\mu(Q=m_{\rm stop})$ generated by only gaugino loops in low scale gauge mediation tends to be inconsistent with the electroweak symmetry breaking and/or the LHC constraints on $m_{H/A}$. Generally, the process that generates a $\mu$ term will generate a $B_\mu$ term unless additional symmetries are imposed which forbid operators like $K=|Z|^2 H_u H_d + h.c.$.

In summary, the generic parameters of the gauge mediation model for explaining $g_\mu -2$ are as follows:\footnote{The previous models shown in Ref.~\cite{Bhattacharyya:2018inr} and \cite{Ibe:2021cvf}, can only reproduce a portion of the parameter space presented here.
	In Ref.~\cite{Bhattacharyya:2018inr}, only the case $m_{H_u}^2=m_{H_d}^2$ (with $B \neq 0$ at the messenger scale) is considered while in Ref.~\cite{Ibe:2021cvf}, the authors only consider the case where the $B$-term is essentially zero at the messenger scale. 
	The differences may seem slight at first glance, but the phenomenological and cosmological 
	implications for gravitino dark matter and the masses of the heavy Higgs bosons can be quite extreme.}
\begin{eqnarray}
M_{\rm mess},\  \Lambda_D\equiv \kappa_D F_Z/M_{D},\ 
r_L\equiv (\kappa_L/M_L)/(\kappa_D/M_{D}),\  \Delta m_{H_u}^2, \ \Delta m_{H_d}^2, \ B_\mu, \ N_5, \label{eq:params_gmsb}
\end{eqnarray}
where $|M_D|=|M_{L}|=M_{\rm mess}$ for simplicity; $N_5$ is the number of the messenger pairs $(\Psi_{\bar D}, \Psi_L)$, which we included solely for generality. The Higg soft mass parameters and $B_\mu$ can be related to $\mu$, $m_A$ and $\tan\beta$ through the electroweak symmetry breaking conditions. This gives us an effective set of parameters which we will use below of 
\begin{eqnarray}
M_{\rm mess},\  \Lambda_D\equiv \kappa_D F_Z/M_{D},\ 
r_L\equiv (\kappa_L/M_L)/(\kappa_D/M_{D}),\  \mu, \ m_A, \ \tan\beta, \ N_5~. \label{eq:params_gmsbmumA}
\end{eqnarray}
Furthermore, since we apply the electroweak symmetry breaking conditions at the weak scale, the boundary conditions on the $\mu$, $m_A$, and $\tan\beta$ are set at the weak scale, and thus $m_A$ corresponds to the physical mass of the pseudo-scalar Higgs boson.

In Eq. (\ref{eq:params_gmsb}), $\Lambda_D$, $r_L$ and $B_\mu$ are in general complex and their phases plus the phase of the $\mu$ parameter cannot be simultaneously removed by any combination of Higgs and messenger field redefinitions and a $U(1)_R$ transformation. Because of this, the predicted electron electric dipole moment (EDM) will easily exceeds the experimental bound~\cite{ACME:2018yjb} unless the phase is drastically suppressed. For the moment, we assume $\Lambda_D$, $r_L$, $B_\mu$ and $\mu$ can be made real. We will revisit this issue at the end of this section.

We now give expressions for the soft masses at the messenger scale, which are 
\begin{eqnarray}
M_{\tilde b} \simeq N_5 \frac{g_1^2}{16\pi^2} \Lambda_D \left(\frac{3}{5} r_L + \frac{2}{5} g(x) \right), \ 
M_{\tilde w} \simeq N_5 \frac{g_2^2}{16\pi^2} r_L \Lambda_D, \   
M_{\tilde g} \simeq N_5 \frac{g_3^2}{16\pi^2} \Lambda_D g(x), \label{eq:gaugino_masses}
\end{eqnarray}
where $g_1$, $g_2$ and $g_3$ are the SM gauge couplings of $U(1)_Y$, $SU(2)$ and $SU(3)$, $x=|\kappa_D F_Z/M_D^2|$ and $g(x)$ is a loop function given in Ref.~\cite{Martin:1996zb},
\begin{eqnarray}
g(x) = \frac{1}{x^2}\left[(1+x)\ln(1+x)\right] + (x \to -x)~,
\end{eqnarray}
and the factor of $r_L$ in the gaugino masses comes from expanding $g(x)$ in the limit $r_L\ll 1$. The soft SUSY breaking masses for sleptons $(\tilde{L}, \tilde{E})$, squarks $(\tilde{Q}, \tilde{U}, \tilde{D})$ and Higgs fields $(H_u, H_d)$ are 
\begin{eqnarray}
m_{\tilde{L}}^2 &\simeq&N_5  \frac{|\Lambda_D|^2}{(16\pi^2)^2}\left(\frac{3}{2}|r_L|^2 g_2^4 + \frac{1}{4}g_1^4 \Lambda_{1,{\rm eff}}^2 \right),\nonumber \\
m_{\tilde{E}}^2 &\simeq&N_5  \frac{|\Lambda_D|^2}{(16\pi^2)^2}\left(g_1^4 \Lambda_{1,{\rm eff}}^2 \right),\nonumber \\
m_{\tilde{Q}}^2 &\simeq&N_5  \frac{|\Lambda_D|^2}{(16\pi^2)^2}\left(\frac{8}{3}g_3^4 f(x) + \frac{3}{2}|r_L|^2 g_2^4 + \frac{1}{36}g_1^4 \Lambda_{1,{\rm eff}}^2 \right),\nonumber \\ 
m_{\tilde{U}}^2 &\simeq&N_5  \frac{|\Lambda_D|^2}{(16\pi^2)^2}\left(\frac{8}{3}g_3^4 f(x)+\frac{4}{9}g_1^4 \Lambda_{1,{\rm eff}}^2\right),\nonumber \\
m_{\tilde{D}}^2 &\simeq&N_5  \frac{|\Lambda_D|^2}{(16\pi^2)^2}\left(\frac{8}{3}g_3^4 f(x)+ \frac{1}{9}g_1^4 \Lambda_{1,{\rm eff}}^2\right),\nonumber \\
m_{H_u}^2 &\simeq& m_L^2 + \Delta m_{H_u}^2, \nonumber \\
m_{H_d}^2 &\simeq& m_L^2 + \Delta m_{H_d}^2, \label{eq:sfermion_masses}
\end{eqnarray}
where
\begin{eqnarray}
\Lambda_{1,{\rm eff}}^2 &=& \frac{6}{5}\left[\frac{3}{5} |r_L|^2 + \frac{2}{5} f(x)\right], \nonumber \\
f(x) &=& \frac{(1+x)}{x^2}\left[ \ln(1+x)-2{\rm Li}_2 \left(\frac{x}{1+x}\right) + \frac{1}{2}{\rm Li}_2\left(\frac{2x}{1+x}\right)\right] + (x \to -x).
\end{eqnarray}
The $|r_L|^2$ in the sfermion masses again comes from expanding $f(x)$ in the limit that $r_L\ll1$

From these expressions, it is clear that light sleptons and electroweak gauginos are obtained for $|r_L|=|\kappa_L/\kappa_D| \ll 1$ while the squarks and gluino remain heavy. It is sometimes important to include the full function $g(x)$ and $f(x)$ for the contributions coming from $\Psi_D$ and $\Psi_{\bar D}$ but it is never important for the $\Psi_L$ and $\Psi_{\bar L}$ contribution since we will always take $|r_L| \ll 1$. We also see from these expressions that as $N_5$ increases the masses of the electroweak gauginos become larger for fixed slepton masses. This is particularly true when $M_{\rm mess}$ is small.\footnote{When $M_{\rm mess}$ is large, this effect is offset by the RG running of the soft masses} If we consider the LHC constraints on slepton masses, it becomes clear that it is marginally easier to explain $g_\mu-2$ for $N_5=1$ than $N_5>1$. For $N_5>1$, we need an even larger $\tan\beta$, which in turn makes the staus lighter as we discuss later. Therefore, we will focus on the $N_5=1$ case in what follows.

\begin{figure}[htp]
\centering
\includegraphics[scale=0.5]{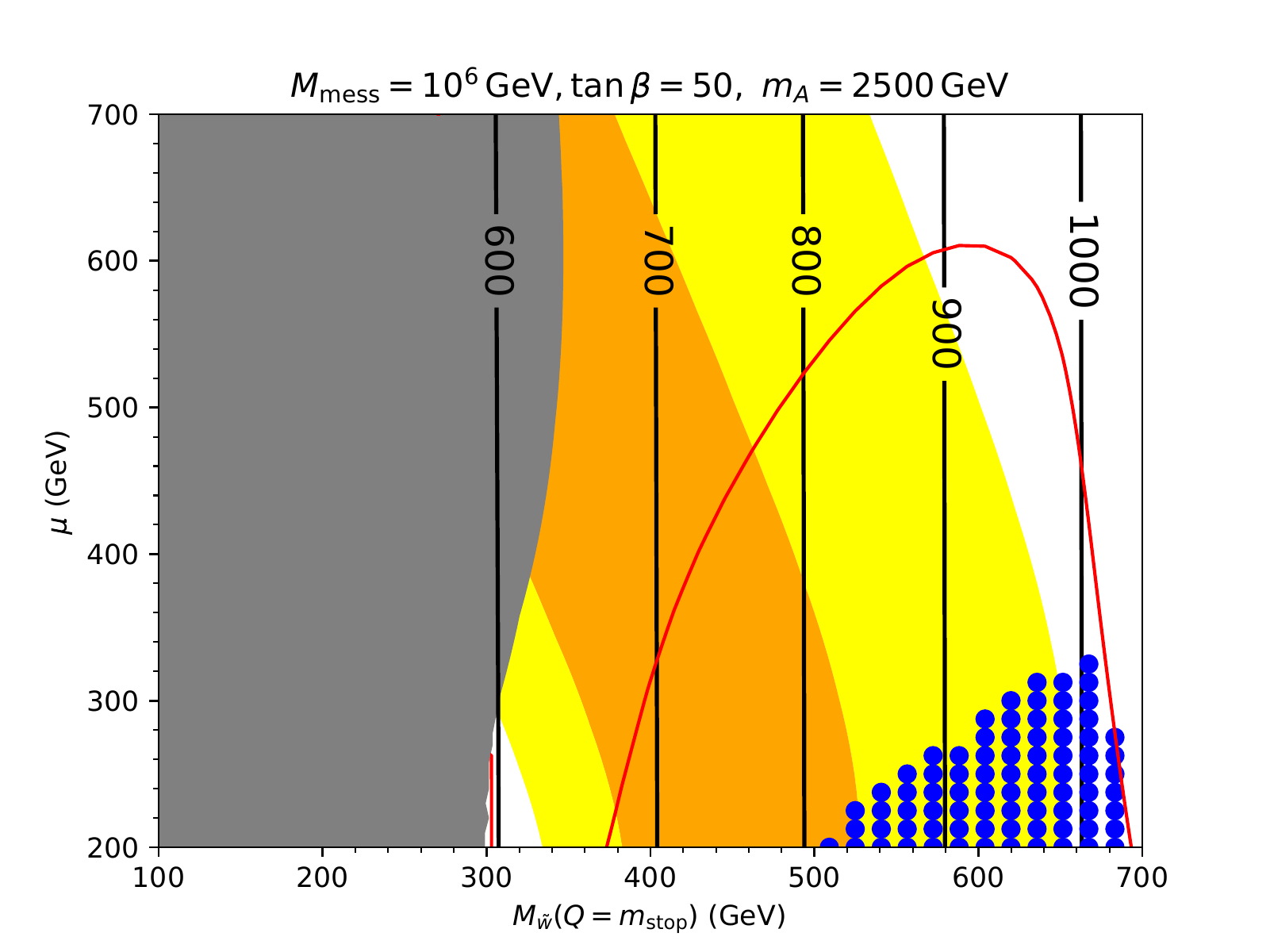}
\includegraphics[scale=0.5]{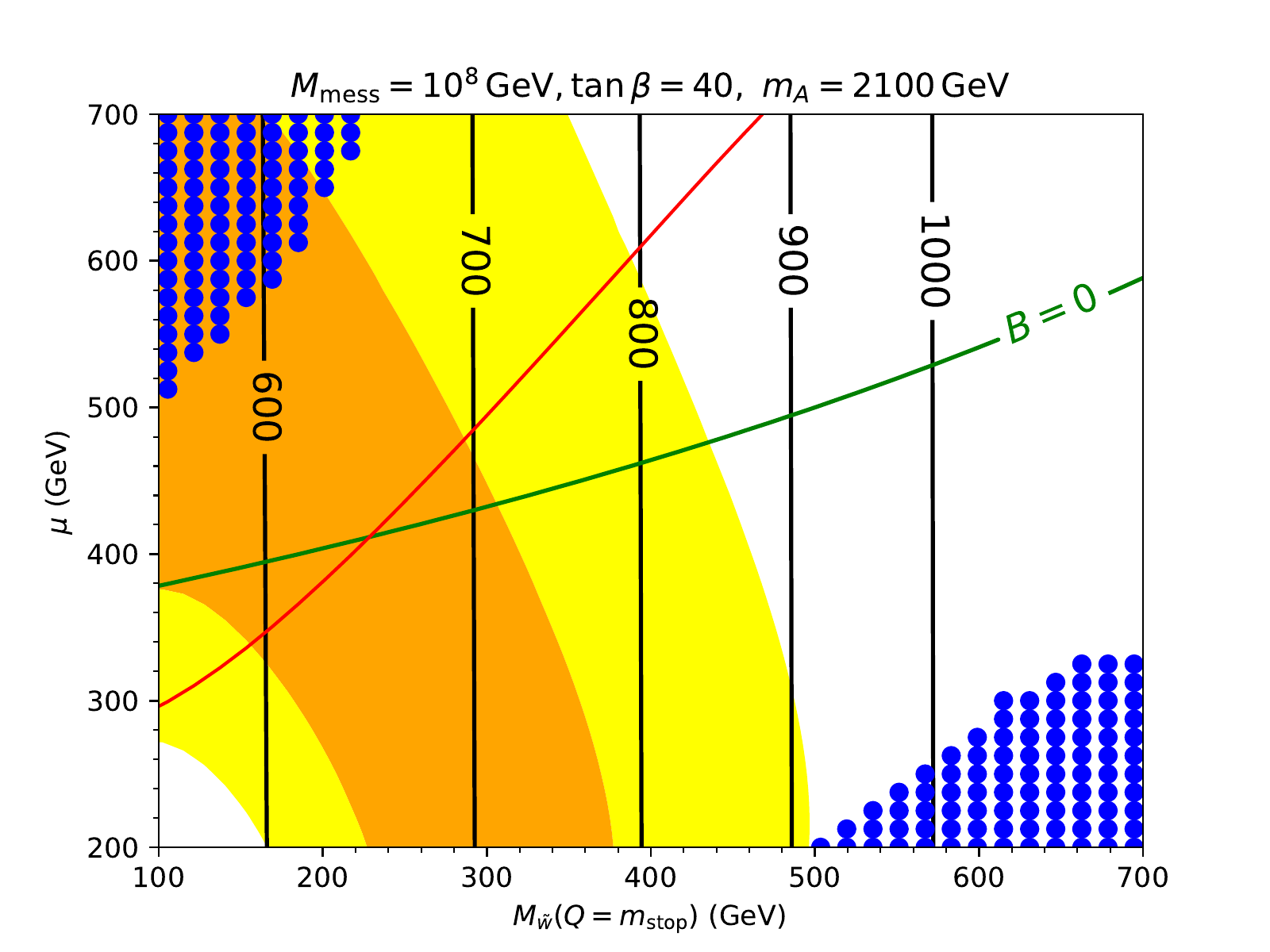}
\includegraphics[scale=0.5]{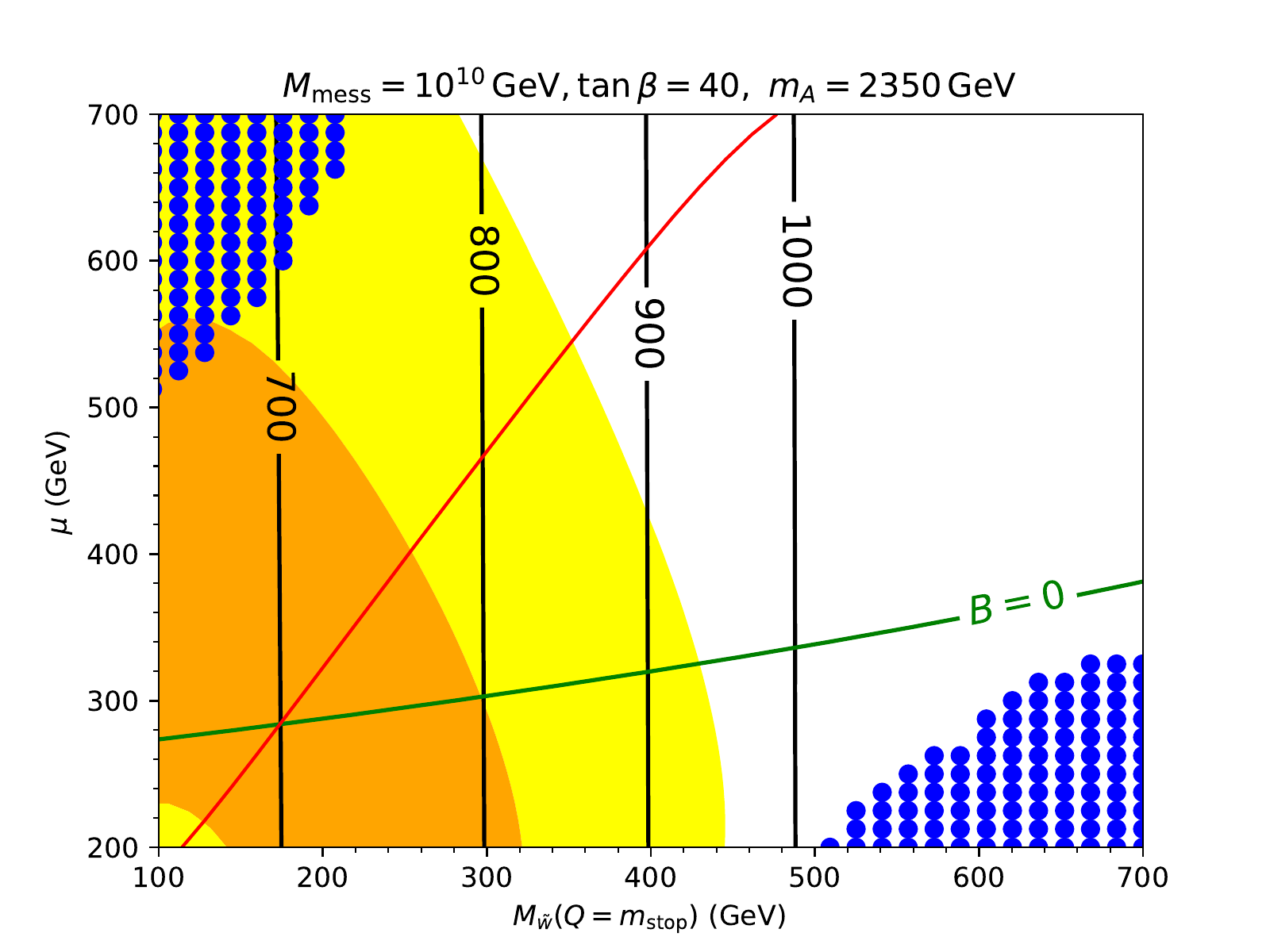}
\includegraphics[scale=0.5]{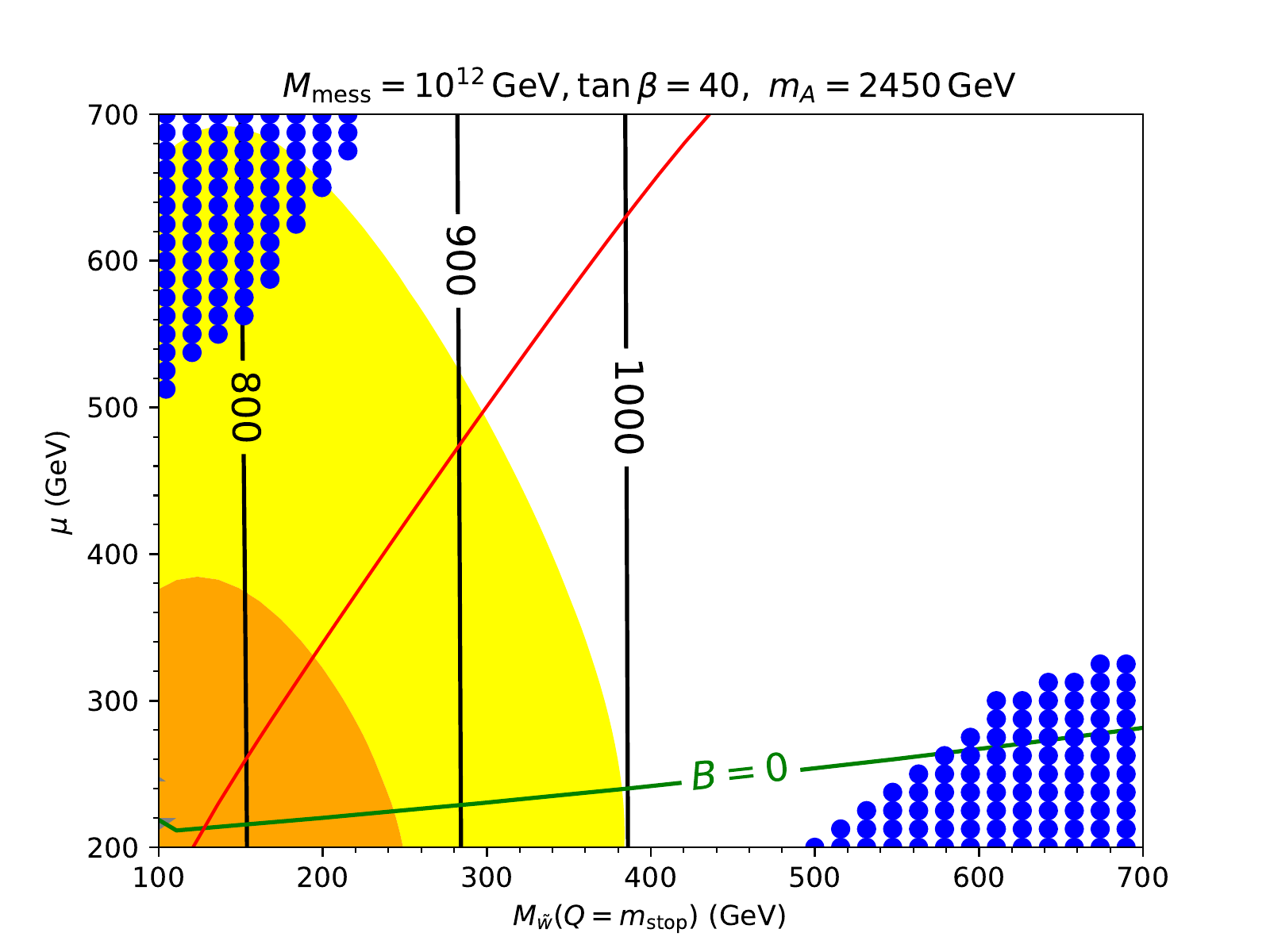}
\caption{The $M_{\tilde w}-\mu$ plane for $\Lambda_D=900$\, TeV. The black lines are contours of $m_{{\tilde \nu}_\mu}$ and the orange (yellow) shaded regions are consistent with $g_\mu -2$ at the 1-$\sigma$ (2-$\sigma$) level. The gray shaded region is excluded due to a tachyonic (or NLSP) stau. In the regions above the red solid lines, $m_{\chi_2^\pm} > m_{\tilde{\tau}_1}$. The blue-dotted regions are excluded due to the LHC constraint in Ref.~\cite{ATLAS:2021yqv}. The green lines correspond to $B_\mu(Q=M_{\rm mess})=0$.}
\label{fig:1}
\end{figure}
\begin{figure}[htp]
	\centering
	\includegraphics[scale=0.45]{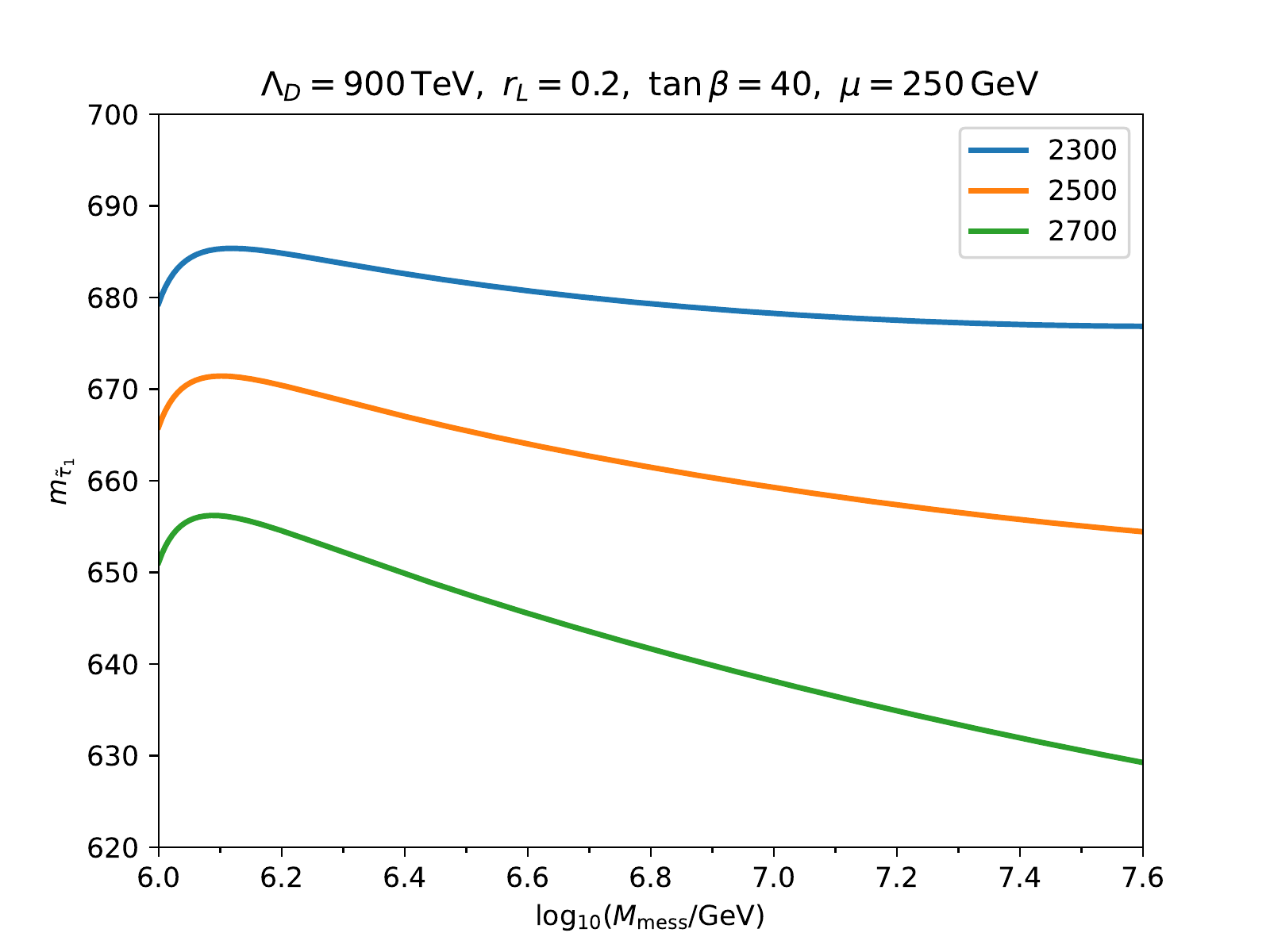}
	\includegraphics[scale=0.45]{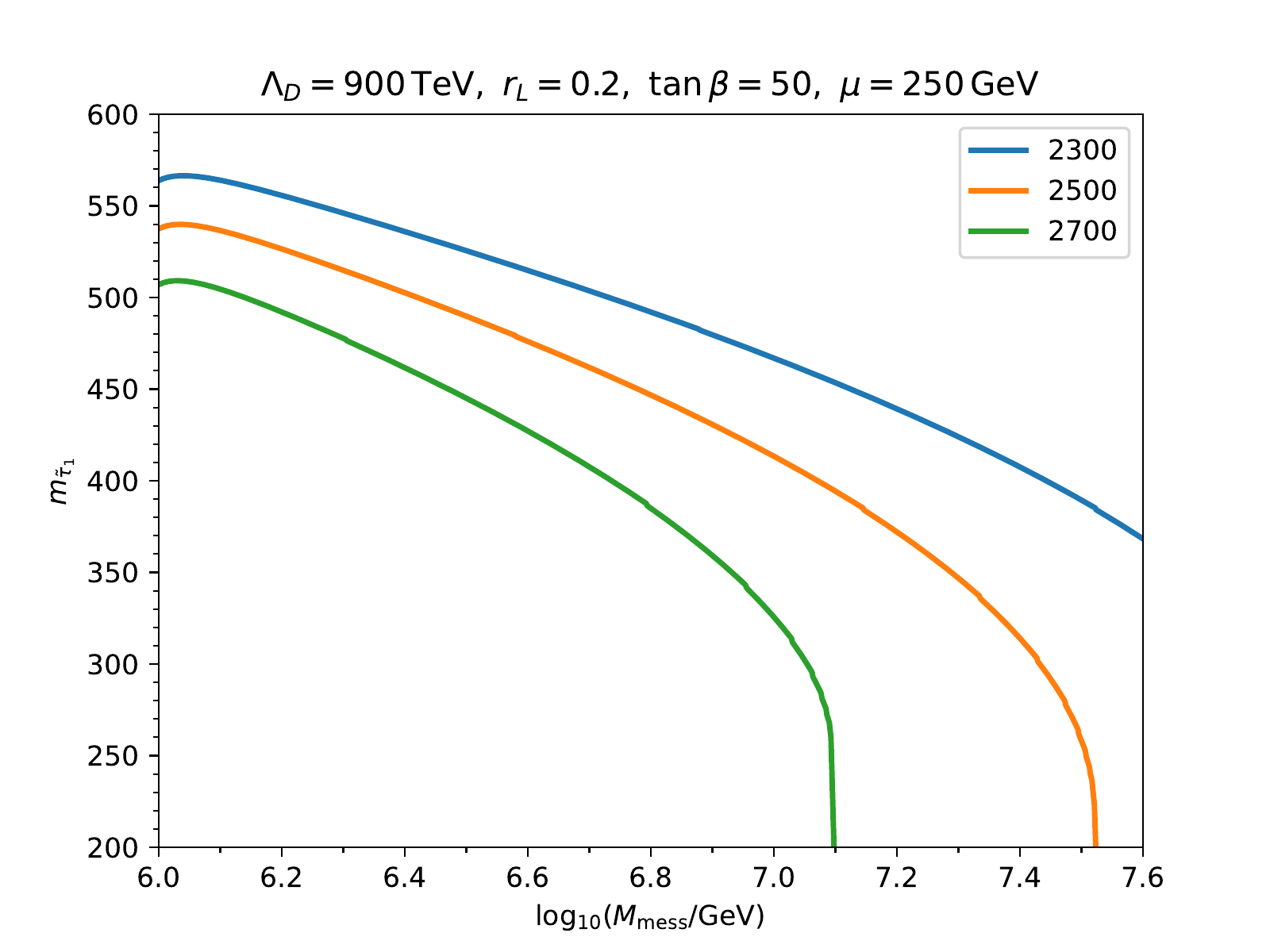}
	\caption{The lightest stau mass as a function of the messenger scale for different $m_A$, $m_A=(2300,2500,2700)$\,GeV. In the left (right) panel, $\tan\beta=40$ (50).}
	\label{fig:mstau}
\end{figure}
\begin{figure}[htp]
\centering
\includegraphics[scale=0.5]{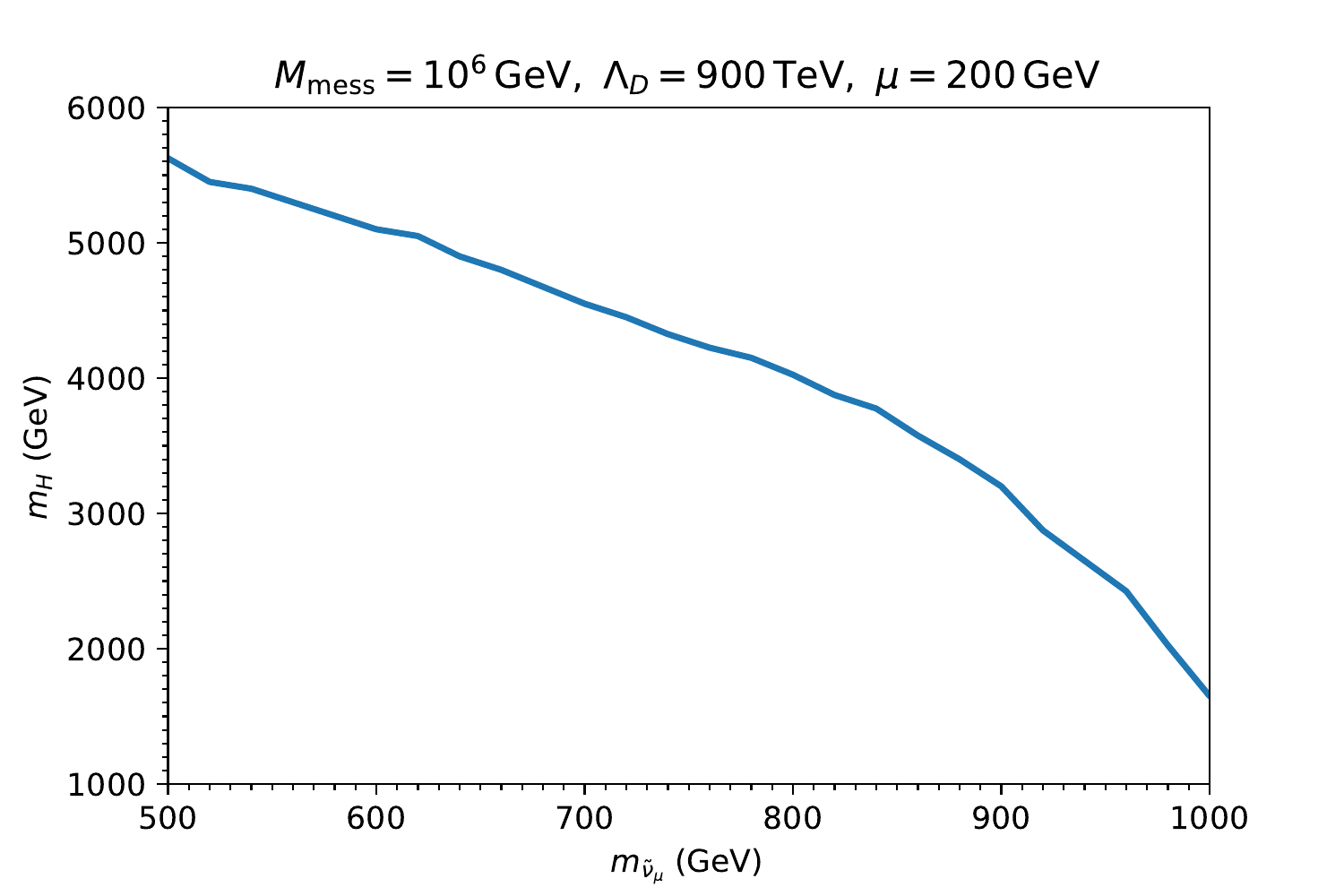}
\includegraphics[scale=0.5]{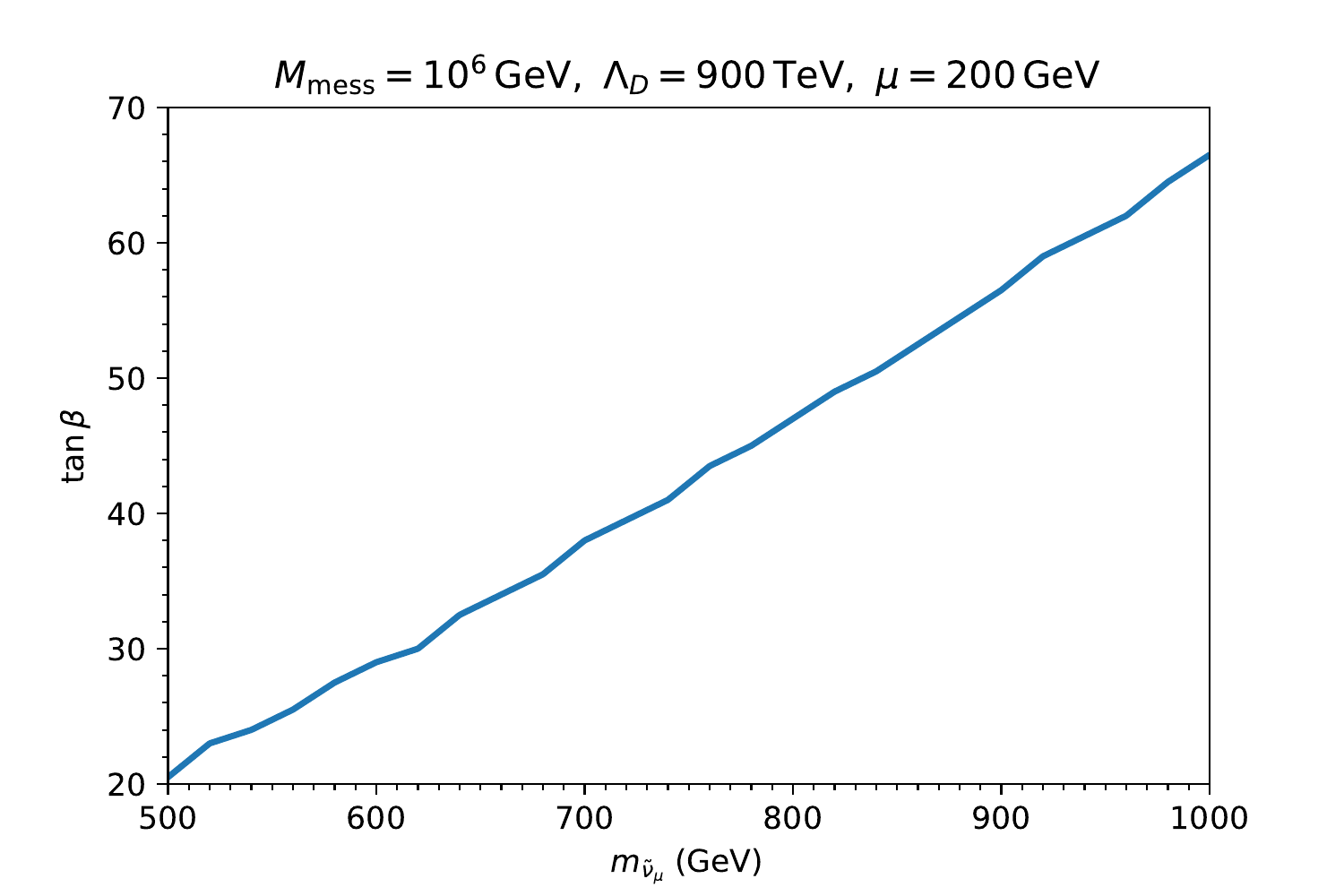}
\includegraphics[scale=0.5]{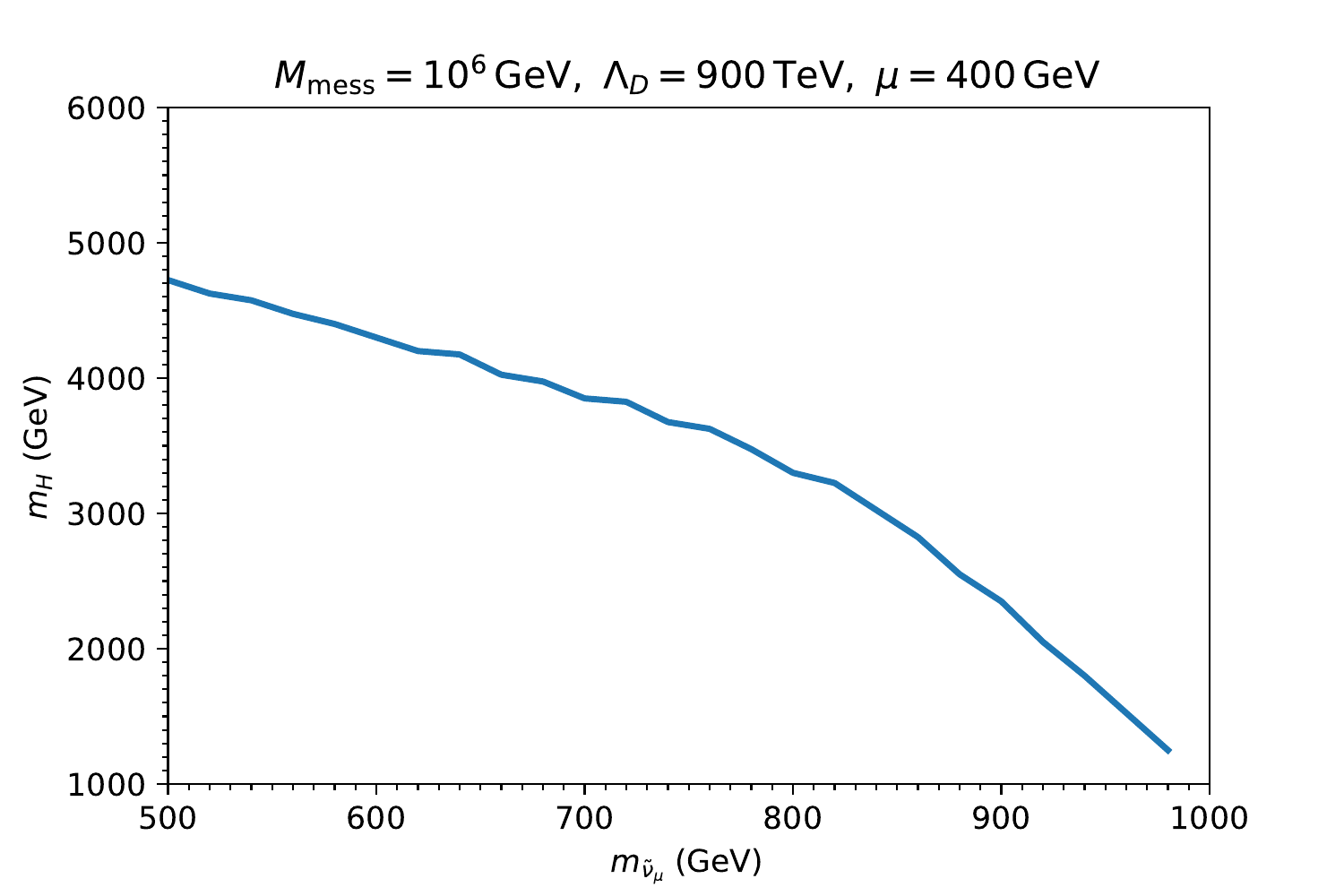}
\includegraphics[scale=0.5]{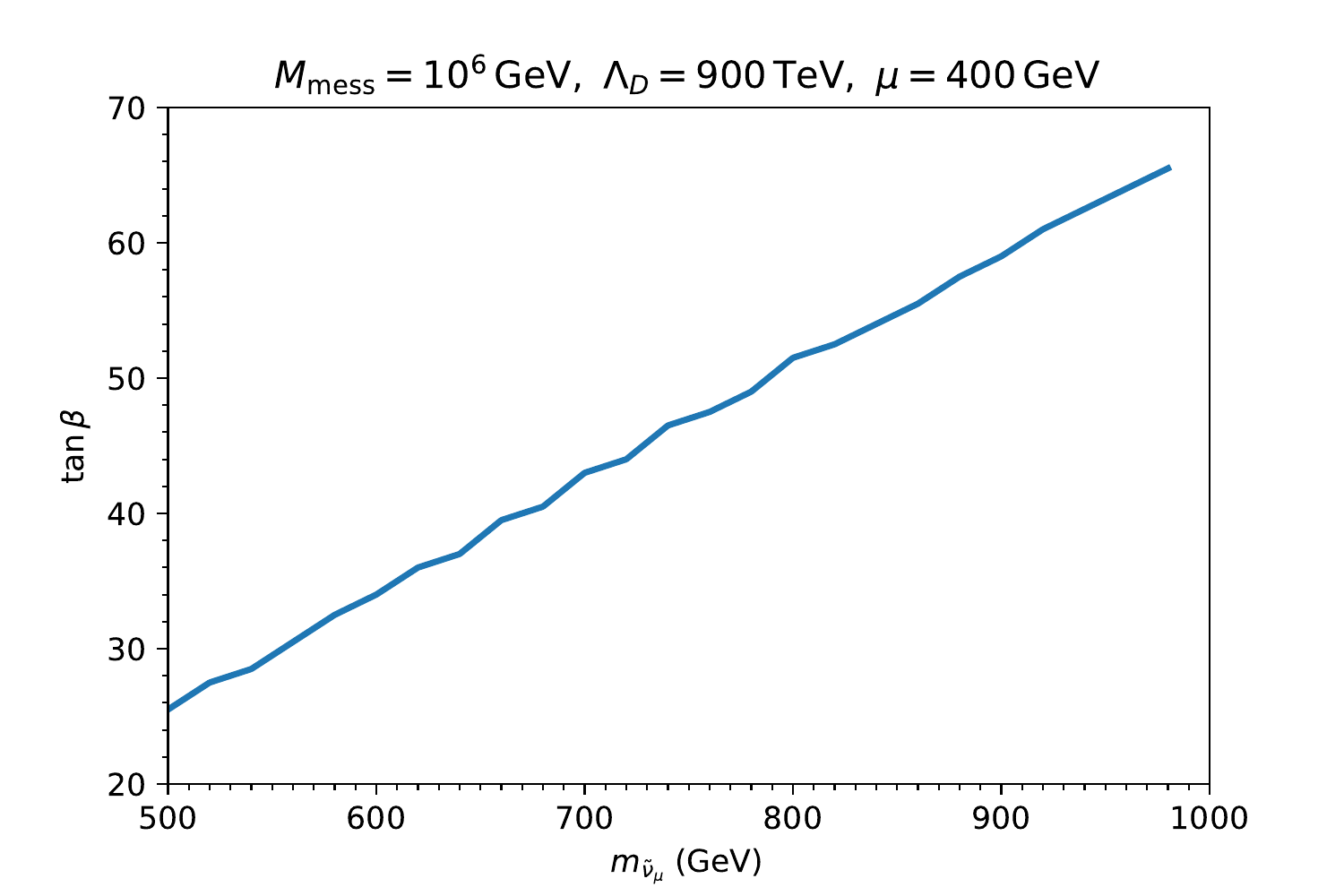}
\caption{The maximum value of $m_{H}$ (left) and corresponding $\tan\beta$ (right) as a function of $m_{\tilde{\nu}_{\mu}}$ when we require $19.2 \leq (a_\mu)_{\rm SUSY}\times 10^{10}$ and the NLSP is a neutralino.}
\label{fig:2}
\end{figure}

In Fig.~\ref{fig:1}, we show regions of parameter space which are consistent with $g_\mu-2$ in the $M_{\tilde w}(Q=m_{\rm stop})-\mu$ plane.\footnote{In the figure, when we calculate $M_{\tilde w}$, the threshold correction to $g_2^2(Q=m_{\rm stop})$ from the higgsino loop is evaluated with $\mu=200$\,GeV. In the range of $\mu$ taken in the figure, the difference between the corrections for $\mu=200$\,GeV and $\mu=700$\,GeV is less than 0.5\%.} We take $\mu>0$ and $r_L>0$ (and $\Lambda_D>0$). Otherwise, the relative signs of $M_{\tilde w}$ and $M_{\tilde b}$ is opposite leading to some cancellation between the chargino and bino-loop contributions. This gives a suppression of about $\sim 10\%$ to the prediction for $g_\mu-2$. 
For these figures, we take $\Lambda_D=900$\,TeV, and $r_L$ and $\mu$ are varied. The other parameters are taken to be $(M_{\rm mess}/{\rm GeV}, \tan\beta, m_A/{\rm GeV})$=$(10^6, 50, 2500)$,$(10^8, 40, 2100)$,$(10^{10},40,2350)$ and $(10^{12},40,2450)$ for the four figures.\footnote{We take these values of $m_A$ so that the $B_\mu=0$ line appears on the figure. The other aspects of these figures are fairly insensitive to these small changes.} 
In the orange (yellow) regions, the $g_\mu-2$ is explained at the 1$\sigma$ (2$\sigma$) level.\footnote{
There are regions where the higgsino and wino are as light as 150-200 GeV. In these regions, the predicted $W$-boson mass can be larger than the SM prediction by  3-6\,MeV~\cite{Marandella:2005wc, Strumia:2022qkt}. The large mass shift recently reported by CDFII is very hard to explain in the present model. This will be checked in future experiments.
}
For $m_{\tilde \nu_{e,\mu}} \gtrsim 660$\,GeV, the LHC constraints on the first and second generation sleptons can be avoided~\cite{Ibe:2021cvf,ATLAS:2019lng,CMS:2020bfa}.  
In the gray region of the top-left panel, the stau is either the NLSP or tachyonic. This region is essentially excluded since the parameter sets where the stau is the NLSP and tachyonic are almost identical. For larger messenger scales, we have regions consistent with $B_\mu(Q=M_{\rm mess})=0$, which are seen as the green lines.  
The blue dotted regions are excluded due to the constraints from LHC chargino/neutralino searches with $W/Z/h$ final states~\cite{ATLAS:2021yqv}. The constraints are applied 
when the the left-handed stau is heavier than the wino. In the case where the wino-like chargino/neutralino decays to the stau or the tau sneutrino, the effective cross section with the $W/Z/h$ final states decreases to about 50\,\%
and the constraints in Ref.~\cite{ATLAS:2021yqv} are not directly applicable anymore.     
The predicted mass of the standard model-like Higgs ($h$) is $\approx$\,124.5\,-\,125.5 GeV in the region where $a_\mu$ is in the $1\sigma$ range.\footnote{For the small $\mu$ case, the negative contribution to $m_h^2$ from the sbottom loop is proportional to $(\mu\tan\beta)^4$ and so suppressed. Therefore, the stop mass required to explain a $125$ GeV Higgs boson mass becomes somewhat smaller.}
In the analysis, we use {\tt SOFTSUSY 4.1.12}~\cite{Allanach:2001kg} to evaluate the SUSY mass spectra, {\tt FeynHiggs 2.18.0}\cite{Heinemeyer:1998yj, Heinemeyer:1998np, Degrassi:2002fi, Frank:2006yh, Hahn:2013ria, Bahl:2016brp, Bahl:2018qog} to calculate the lightest Higgs boson mass,
and {\tt GM2Calc 2.1.0}~\cite{Athron:2015rva, Athron:2021evk} to evaluate the SUSY contribution to the $g_\mu-2$.

Now, we discuss the upper bound on $m_{H/A}$. A meaningful upper bound is obtained when the sleptons are heavier, since explaining the $g_\mu-2$ then requires $\tan\beta$ to be pushed to larger values. The upper bound on $m_{H/A}$ then comes from the requirement that the lightest stau is not the next-to-lightest SUSY particle (NSLP) or tachyonic.\footnote{These are effectively the same upper bound.}  Although the vacuum stability constraint from the stau-Higgs potential can be avoided easily in the small $\mu$ case, the soft SUSY breaking masses for the staus still give a meaningful constraint since the stau becomes the NSLP or even tachyonic when $\tan\beta$ and/or $m_{H_d}^2$ are large. This can be understood by examining the renormalizaton group equations:
\begin{eqnarray}
\frac{m_{H_d}^2}{d \ln Q} &\ni& \frac{1}{16\pi^2} \left[6(m_{Q_3}^2 + m_{D_3}^2  + m_{H_d}^2) Y_b^2 + 2 m_{H_d}^2 Y_\tau^2 \right] \nonumber \\
\frac{m_{L_3}^2}{d \ln Q} &\ni& \frac{1}{16\pi^2} (2 m_{H_d}^2 Y_\tau^2) \nonumber \\
\frac{m_{E_3}^2}{d \ln Q} &\ni& \frac{1}{16\pi^2} (4 m_{H_d}^2 Y_\tau^2),
\end{eqnarray}
where $Y_b$ and $Y_{\tau}$ are the bottom and tau Yukawa couplings, respectively; $m_{Q_3}$ and $m_{D_3}$ ($m_{L_3}$ and $m_{E_3}$) are left and right handed sbottom (stau) masses. From the first equation we see that $m_{H_d}^2(Q=M_{\rm mess})$ must be much larger than $m_{H_d}^2(Q=m_{\rm stop})$ as it receives large corrections from sbottom loops.\footnote{Because the weak gauginos are light, they have a subdominat effect.} This is particularly true when $\tan\beta$ is large since $Y_b$ and $Y_\tau$ grow with $\tan\beta$. Since the stau masses receive corrections proportional to $|Y_\tau|^2m_{H_d}^2$, they tend to be suppressed as the messenger scale is increased. 
In Fig.~\ref{fig:mstau}, the lightest stau mass as a function of the messenger scale is shown. The stau mass monotonically decreases as the messenger scale increases.
For large $m_{H/A} \sim m_{H_d}(Q=m_{\rm stop})$ and large $\tan\beta$, the staus mass will be too small or tachyonic. As the slepton masses are increased, the required $\tan\beta$ for explaining $g_\mu-2$ becomes larger leading to an upper bound on $m_{H/A}$. Since the mass suppression discussed above is definitely smaller for the lower messenger scales, we will take a low messenger scale as a reference point. (The complications are much more severe for larger SUSY breaking mediation scales like the Planck or GUT scales. In this sense, gauge mediation is advantageous for explaining $g_\mu-2$ via the chargino loop contribution.)

In the left panel of Fig.\,\ref{fig:2}, we show the upper bound on $m_{H}(\simeq m_A)$ as a function of the muon sneutrino mass, $m_{\tilde{\nu}_\mu}$. The right panel shows the value of $\tan\beta$ for the maximal value of $m_{H}$. In these figures, we scan the following parameter ranges: 
\begin{eqnarray}
0.1 \leq r_L \leq 0.3, \ 20\leq \tan\beta \leq 70, \ 1000 \leq m_A/{\rm GeV}\leq 6000.
\end{eqnarray}
We require that $(a_\mu)_{\rm SUSY}$ explain the experimental value of the $g_\mu-2$ at 1 $\sigma$ level and the NLSP is a neutralino. The messenger scale and $\mu$ are fixed to be $M_{\rm mess}=10^6\,{\rm GeV}$ and $\mu=200$\,GeV so that the chargino contribution to the $g_\mu-2$ is (almost) maximized and the negative contribution to the square of the stau masses from renormalization group running is minimized. We also take $\Lambda_D=900\,{\rm TeV}$. This shows that for larger $m_{{\tilde \nu}_\mu}$, the maximum value of $m_{H/A}$ becomes smaller while $\tan\beta$ becomes larger. In particular, for $m_{{\tilde \nu}_\mu}\gtrsim 900$\,GeV, we get $m_{H/A} \leq 3$\,TeV and $\tan\beta \gtrsim 56$. This shows the possibility to detect the heavy Higgs bosons, $H$ and $A$,  at the LHC, using the channel $H/A \to \tau \tau$ is quite good since $Y_b$ and $Y_{\tau}$ are quite large.

In table~\ref{tb:masses}, we show the mass spectra, $(a_\mu)_{\rm SUSY}$, the life-time of the lightest chargino, and production cross section ($bbH$) of $H/A$ at $\sqrt{s}=14\, {\rm TeV}$ and branching ratios of $H/A$ for four different mass spectra. We also show the required $\Delta m_{H_u}^2$, $\Delta m_{H_d}^2$ and $B_\mu$ at the messenger scale. The production cross sections are calculated using {\tt SusHi 1.7.0} package~\cite{Harlander:2012pb,Harlander:2016hcx} with {\tt NNPDF4.0}~\cite{NNPDF:2021uiq} used for the parton distribution functions. At the point P4, the production cross section of $H/A$ is quite large. In fact, P4 is marginally consistent with the current bound from the heavy Higgs search~\cite{ATLAS:2020zms} and it is expected that the point will be checked soon~\cite{ATL-PHYS-PUB-2018-050}. 

Concerning the constraints from the chargino/neutralino searches at the LHC, for P2 and P4, with the left-handed stau lighter than the wino, about 26-27$\%$of wino-like charginos and neutralinos ($\chi_2^\pm$) decay to the stau or tau sneutrino. Therefore, the exclusion limits using the $W/Z/h$ final states become weaker as the effective cross section decreases to about 50$\%$. Note, we cannot directly apply the exclusion limit in Ref.~\cite{ATLAS:2021yqv}. In order to check the LHC constraints, including those in Ref.~\cite{ATLAS:2021yqv}, we used {\tt SModelS} package~\cite{Kraml:2013mwa,Ambrogi:2017neo,Dutta:2018ioj,Heisig:2018kfq,Ambrogi:2018ujg,Khosa:2020zar,Alguero:2020grj,Alguero:2021dig} with the wino production cross sections 
at NLO+NLL estimated using {\tt resummino}~\cite{Fuks:2013vua,Debove:2009ia,Debove:2010kf,Debove:2011xj,Fuks:2012qx,Fiaschi:2018hgm,Fiaschi:2020udf}. We found P2 and P4 were not excluded.

\begin{table*}[t]
\caption{The mass spectra and $(a_\mu)_{\rm SUSY}$. The (lightest) squark and gluino masses as well as $\Lambda_D$ are shown in units of TeV while the others are shown in units of GeV. We also show $\sqrt{\Delta m_{H_u}^2}$, $\sqrt{\Delta m_{H_d}^2}$ and $\sqrt{B_\mu}$ at the messenger scale.}
\begin{center}
\begin{tabular}{c c c c c c c}
& P1 & P2 & P3 & P4 \\  
\hline
$\Lambda_D$              & 900 & 980 & 950 & 980 \\
$r_L$                              & 0.17 & 0.25 & 0.14 & 0.26\\
$M_{\rm mess}$          & $10^6$ & $10^6$ & $10^8$ & $1.3 \times 10^6$\\
$\mu$                          & 210 & 260 & 441 & 250\\
$m_A({\rm pole})$     & 2500 & 2500 & 2300 & 2000\\
$\tan\beta$               & 40 & 60 & 43 & 64.5\\
\hline 
$\tilde{g}$                & 7.0 & 8.2 & 6.2  & 7.1\\
$\tilde{q}$                & 8.3 & 9.2 & 8.0  & 8.9 \\
${\tilde{\tau}_1}$  & 581 & 300 & 459  & 280 \\
${\tilde{\tau}_2}$  & 830 & 584 & 760  & 583 \\
$\tilde{\mu}_{R}$   & 985 & 1091  & 1118 & 1177 \\
$\tilde{\nu}_{\mu}$ & 687 & 935 & 736  & 987 \\
$({\chi^0_1},{\chi^\pm_1})$ 
                    & (200.2, 206.4) & (257.6, 262.5) 
                    & (322.3, 323.3) & (247.2, 252.3) \\
${\chi^\pm_2}$      & 427 & 667 & 480  & 694\\
${\chi^0_2}$        & 223 & 271 & 456  & 260 \\
${\chi^0_3}$        & 426 & 666 & 471 &  694 \\
${\chi^0_4}$        & 735 & 923 & 636 &  823\\
\hline
$m_h$ & 125.2 & 125.2 & 124.8 & 125.0 \\
$(a_\mu)_{\rm SUSY}/10^{-10}$ & 22.7 & 19.3 & 19.8 & 19.7 \\
$\tau_{\chi_1^\pm}$(ns) 
& $5.7 \times 10^{-8}$ & $2.1 \times 10^{-7}$ 
& $1.8 \times 10^{-4}$ & $1.7 \times 10^{-7}$ \\
$\sigma_{H/A}/10^{-4}$(pb) & 8.1 & 18.0 & 17.4 & 110.3 \\ 
Br$(H \to \tau {\bar \tau})$ & 0.10 & 0.14 & 0.12 & 0.15 \\ 
Br$(H \to b {\bar b})$ & 0.53 & 0.66 & 0.56  & 0.70 \\ 
\hline 
$\sqrt{\Delta m_{H_u}^2}$ & 3887 & 4139 & 4878  & 4137\\
$\sqrt{\Delta m_{H_d}^2}$ & 3480 & 4946 & 3949  & 5084\\
$\sqrt{B_\mu}$ & 385 & 274 & $\approx 0$ &        171 \\
\end{tabular}
\end{center}
\label{tb:masses}
\end{table*}%
Lastly, we discuss the problem of CP violation. Since the smuons and electroweak gauginos need to be as light as $\mathcal{O}$(100)\,GeV to explain the $g_\mu-2$ and we consider gauge mediation, the selectrons must be light as well. The electron (EDM) constraints then require $B_\mu, r_L$ and $\Lambda_D$ to be almost completely real. 

To analyze the electron EDM, we rotate the fields so that $B_{\mu}(Q)$ is real and $\left<H_u^0\right>$ and $\left<H_d^0\right>$ are in turn real and positive. We then take  $\mu=|\mu| \exp(-i\theta_{B})$. In this basis, the potentially problematic phases are:\footnote{Although the complex argument of the scalar trilinear coupling, the $A$-term, is also relevant, we focus on the above mentioned arguments as the contribution to the EDM from picking up the $A$-term is not enhanced by $\tan\beta$.}
\begin{eqnarray}
\theta_{\tilde w} &=& {\rm Arg}(M_{\tilde w} |\mu|\exp(-i\theta_{B})) = {\rm Arg}(M_{\tilde w}  (B_\mu/\mu)^*), \nonumber \\
\theta_{\tilde b} &=& {\rm Arg}(M_{\tilde b} (B_\mu/\mu)^*),
\end{eqnarray}
where all relevant parameters are defined at the low energy scale. The phases of these mass combinations are important because the same diagrams with the same combination of masses contribute to $g_\mu-2$ and the electron (EDM) calculation.\footnote{The one-loop SUSY contribution to $a_e$, the anomalous magnetic moment of the electron, is $(m_e/m_u)^2$ times $a_\mu$ as the selectron and smuon masses are nearly degenerate.} As is the case for the $g_\mu-2$, the chargino diagram gives the largest contribution to the electron EDM and $\theta_{\tilde w}$ is most strongly constrained. 

Furthermore, 
since $a_e$ and the electric dipole moment of the electron, $d_e$, come from the real and imaginary parts of the same amplitude, $\tan({\rm Arg}({\rm Amplitude})) a_e$ is equal to $ 2m_e d_e/e$. Combining these relations, we find
\begin{eqnarray}
(d_e/e)_{\rm SUSY} &\simeq& \tan\theta_{\tilde w}\frac{m_e}{2m_\mu^2} (a_{\mu})_{\rm SUSY} \nonumber \\
&\approx& \theta_{\tilde w} \left(\frac{(a_{\mu})_{\rm SUSY}}{2.2 \times 10^{-9}}\right) \times 10^{-24}\, {\rm cm},
\end{eqnarray}
where we used the fact that the chargino contribution dominates in our model.\footnote{The phase $\theta_{\tilde{b}}$ is also constrained by the electron EDM measurement, but to a lesser degree. However, the model we give below will suppress all problematic phases.} 

In order to satisfy the ACME experimental constraint~\cite{ACME:2018yjb}, $|d_e/e|<1.1 \times 10^{-29}\,{\rm cm}$, $\theta_{\tilde w}$ should be smaller than about $10^{-5}$. The constraint on $\theta_{\tilde b}$ is somewhat weaker, $\theta_{\tilde b} \leq \mathcal{O}(10^{-4})$, since the bino loop contribution only makes up  $\mathcal{O}(10\%)$ of the anomalous magnetic moment calculation. It is extremely difficult to achieve such a small CP violating phase unless the complex arguments of $\Lambda_D$, $r_L \Lambda_D$ and $B_\mu/\mu$ are identical to a very high level of precision. The requirement on $\Lambda_D$ is due to the fact that the complex arguments of $M_{\tilde b}$, $M_{\tilde w}$, $M_{\tilde g}$ are mixed through two-loop RG running. Thus, the phases of all gaugino masses feed into the phase of $B_\mu/\mu$ from RG running. This leaves only one possible conclusion; the complex arguments of $M_{\tilde b}$, $M_{\tilde w}$, $M_{\tilde g}$ and $B_\mu/\mu$ must be aligned at some higher energy scale, i.e. the messenger scale.

To estimate this effect, we can approximate the Higgs $B$-term  at the low scale as~\cite{Martin:1993zk} 
\begin{eqnarray}
B_\mu(Q)/\mu(Q) &\approx& (B_\mu/\mu) - \frac{1}{16\pi^2} \left(6 g_2^2 M_{\tilde w} + \frac{6}{5}g_1^2 M_{\tilde b} \right)\ln\frac{M_{\rm mess}}{Q} \nonumber \\
&+& \frac{1}{8\pi^4} g_3^2 (Y_t^2 + Y_b^2) M_{\tilde g} \left[\ln\frac{M_{\rm mess}}{Q}\right]^2, \label{eq:b-term}
\end{eqnarray}
as long as the messenger scale is not too high. Also, we have taken $Q=m_{\rm stop}$. Although the gluino mass contribution is loop suppressed compared to the wino and bino mass contribution, the loop suppression of the gluino contribution is offset by a larger mass and larger couplings and so can be comparable or even larger than the bino and wino contributions in the parameter space of interest. From Eq.~(\ref{eq:b-term}) it becomes clear, $B_\mu(Q)/\mu(Q)$ picks up a portion of all the complex arguments of the gaugino masses. This, combined with the RG mixing of the gaugino masses, tells us we need a model that predicts the complex arguments of $\Lambda_D$, $r_L\Lambda_D$ and $B_\mu/\mu$ are aligned. How to achieve the needed alignment/suppression of the phases of  $\Lambda_D$, $r_L\Lambda_D$, $B_\mu$ and $\mu$  will be explained in the next section.

\section{A CP-safe gauge mediation model for the $g-2$ anomaly}
In this section, we give an explicit model, which explains $\theta_{\tilde w}=\theta_{\tilde b}=0$, after the gravitino mass is rotated to be real and positive, $m_{3/2} \in \mathbb{R}_{>0}$. Our model is based on a discrete symmetry, $Z(3)_1 \times Z(3)_2\times Z(3)_3 \times Z_{4R}$ with the following superpotential interactions:
\begin{equation}
W=\frac{\lambda_i}{3} S_i^3 + k_i S_i \Psi_i{\bar \Psi_i}   (i=1,2,3). \label{eq:esy1}
\end{equation}
We also introduce tachyonic soft SUSY breaking masses for $S_i$ of
\begin{equation}
V_{\rm soft}= - m^2_{S_i} |S_i|^2. \label{eq:esy2}
\end{equation}
An example model for generating $m^2_{S_i}$ is given in Appendix \ref{sec:a1}. Here, $\Psi_1=H_u$ (${\bar \Psi_1}=H_d$), $\Psi_2 =\Psi_{\bar L}, \Psi_3=\Psi_{\bar D}$, 
where $\Psi_{\bar D}$ and $\Psi_{L}$ are the anti-quark and lepton-like messengers with charge assignments of a $5^*$ representation in an $SU(5)_{\rm GUT}$ theory. However, we do not assume grand unification in this paper. The charges of $S_i$, $\Psi_i$ and ${\bar \Psi}_i$ under $Z(3)_1 \times Z(3)_2\times Z(3)_3 \times Z_{4R}$ are given in Table \ref{tb:charges} in Appendix \ref{sec:a1}. The discrete $R$ symmetry, $Z_{4R}$, is anomaly free with one pair of $\Psi_{2,3}$ and $\bar{\Psi}_{2,3}$~\cite{Kurosawa:2001iq}. Therefore it could be a gauge symmetry and left unbroken even with gravitational effects~\cite{Krauss:1988zc, Preskill:1990bm, Banks:1991xj, Ibanez:1991hv, Ibanez:1991pr, Ibanez:1992ji} included. 
The $\lambda_i$ can be made real and positive by a field redefinition of the $S_i$ without any loss of generality. Using a $U(1)_R$ rotation, we take the gravitino mass, $m_{3/2}=\left<W\right>^*/M_{\rm PL}^2$, to be real and positive.
Due to the tachyonic mass, both the scalar and $F$ component of the superfield $S_i$ obtain a non-zero VEV, giving the SUSY preserving and SUSY breaking masses for the messenger and Higgs fields. Note that, as shown in the previous section, there are some regions of parameter space which can explain $g_\mu-2$ and the Higgs $B$-term vanishes at the messenger scale. In this case, we only introduce a higgsino mass term, $\mu H_u H_d$, instead of $S_1$, and the $Z_{4R}$ charge of $H_u H_d$ should be 2. Since this modification is straightforward, we do not include any details here.

Eqs.~(\ref{eq:esy1}) and (\ref{eq:esy2}) give the following tree-level potential for $S_i$: 
\begin{equation}
V = \lambda_i^2 |S_i|^4 - |m^2_{S_i}| |S_i|^2,
\end{equation}
with a minimum at $\left<|S_i|\right>=\sqrt{|m^2_{S_i}|}/\sqrt{2 {\lambda_i}^2}$. 

The phase direction of the $\left<S_i\right>$ are not determined by the tree-level potential and are only determined once loop corrections are taken into account. At the tree level, the scalar potential for $S_i$ in the framework of supergravity is given by
\begin{eqnarray}
V = F_{S_i}^\dag F_{S_i}-3|W|^2/M_{\rm PL}^2, \label{eq:scalar_potential}
\end{eqnarray}
where 
\begin{eqnarray}
F_{S_i} = - \left[ 
\lambda_i S_i^{* 2} + S_i W^*/M_{\rm PL}^2
\right], \label{eq:fsi}
\end{eqnarray}
and we have assumed that the SUSY breaking field has a negligibly small scalar vacuum expectation value. We will return to this point later. As is clear from these expressions, the scalar potential in Eq.\,(\ref{eq:scalar_potential}) does not determine ${\rm Arg}(S_i)$.

The phases of the vevs are determined by the $A$-term contributions to the potential induced by anomaly mediated SUSY breaking (AMSB),
\begin{eqnarray}
V = (A_i)_{\rm amsb} \frac{\lambda_i}{3} S_i^3 + h.c. = \frac{m_{3/2}}{16\pi^2} (16\pi^2 \beta_{\lambda_i}/\lambda_i) \frac{\lambda_i}{3} S_i^3 + h.c.,  \label{eq:aterm_amsb}
\end{eqnarray} 
where $\beta_{\lambda_i}$ is the beta-function for $\lambda_i$, given by
\begin{equation}
(16\pi^2) \beta_i/\lambda_i = 3(N_c |k_i|^2 + 2\lambda_i^2), 
\end{equation}
and $N_c=2$ for $i=1,2$ and $N_c=3$ for $i=3$. 
In the previous paper~\cite{Evans:2010ru}, 
the effects from anomaly mediation were not considered.~\footnote{In the previous paper, the authors introduced higher dimensional operators, $K \ni |S_i|^4/\Lambda_4^2$, to fix the phase direction of the $S_i$. If the cut-off $\Lambda_4$ is larger than $\sim 10 \left< S_i \right>$, the anomaly mediation effects are dominant.} For $S_1$, the tree-level contribution, $(A_1)_{\rm tree}$, shown in Appendix~\ref{ap:treeB} may also be considered, which dominates over the contribution from anomaly mediation $(A_1)_{\rm amsb}$ and has the opposite sign.
By writing $\left< S_i \right> = |\left<S_i\right>| e^{i \theta_{S_i}}$ ($\theta_{S_i}= {\rm Arg}(\left<S_i\right>)$), in Eq.~\eqref{eq:aterm_amsb} (and Eq.~\eqref{eq:a1tree}), $\theta_{S_i}$ has a potential 
\begin{eqnarray}
	V \ni A_i \frac{\lambda_i}{3} S_i^3 + h.c. \to  \frac{2}{3} A_i \left|\left< S_i \right>\right|^3 \lambda_i \cos (3 \theta_{S_i}),
\end{eqnarray}
where $A_i$ corresponds to $(A_{2,3})_{\rm amsb}$ or $(A_{1})_{\rm amsb} + (A_1)_{\rm tree}$, and is a real number. 
The above potential fixes $3 \theta_{S_i} = \pi$ ($3 \theta_{S_i} = 0$) for $A_i>0$ ($A_i<0$).
Since the SUSY masses for $\Psi_i$ are $k_i \left<S_i\right>$ and the $B$-terms are $k_1 \lambda_1 \left<S_1^{*2}\right>$ for $i=1$ and $-k_i \lambda_i \left<S_i^{*2}\right>$ for $i=2$ and $3$,\footnote{%
Following the standard convention, the definitions of the $B$-terms are $W=(\mu - B_\mu \theta^2)H_u H_d$ for the Higgs fields and $W=((M_{\rm mess})_{2,3} + B_{2,3} \theta^2) \Psi_{2,3} \bar{\Psi}_{2,3}$ for the messenger fields. Note that for $\lambda_1 \sim k_1 \sim 1$, $S_1$ significantly mixes with the Higgs doublets after electroweak symmetry breaking. Therefore, it affects the electroweak symmetry breaking as well as the collider and cosmological/astrophysical phenomena, which should be carefully investigated (see e.g., Refs.~\cite{Sabatta:2019nfg, Beck:2021xsv}). On the other hand, for $\lambda_1, k_1 \ll 1$, $S_1$ decouples from the standard model, and the effects are limited.
} we have
\begin{eqnarray}
B_\mu/\mu &=& \pm \lambda_1 |\left<S_1\right>|, \nonumber \\
\Lambda_L (=r_L \Lambda_D) &=& \lambda_2 |\left<S_2\right>|, \nonumber \\
\Lambda_D &=& \lambda_3 |\left<S_3\right>|,
\end{eqnarray}
where $B_\mu/\mu$ is positive (negative) when the tree level contribution (AMSB contribution) to $S_1^3$ term dominates. All of the above parameters are real. Therefore, there is no CP violating phase.

However, the above mechanism is not viable generically. This is because, as shown in the previous section, the phases of the $A_i$'s should be aligned at the order of $10^{-5}$. The required conditions are summarized as follows: 
\begin{itemize}
    \item For SUSY breaking fields which are charged under some symmetry,\footnote{SUSY breaking field means any field whose $F$ component has a non-zero vacuum expectation value.} $|\left<Z_i^*F_{Z_i}\right>/M_{\rm PL}^2| \ll 10^{-7} |m_{3/2}|$.

\item For moduli type SUSY breaking fields (e.g. axion) where the shift symmetry breaking is small, mixings with other SUSY breaking fields could be problematic. This effect is safe if the previous condition is satisfied.

\item There are no true singlets, i.e. no charge under any symmetry, SUSY breaking fields since they generically will have an $F$-term of $O(m_{3/2} M_{\rm PL})$ and will couple to $S_i$, as well as other fields, via a dimension 5 operator with a complex coefficient.

\end{itemize}
To derive these conditions, we have assumed that higher dimensional operators containing the SUSY breaking fields, matter fields and/or moduli fields are suppressed by the Planck scale. If these operators are suppressed by a lower mass scales than the Planck scale, the conditions of above become even more restrictive. For a class of motivated models, we have checked that our CP-safe mechanism still holds (see Appendix \ref{ap:ext}).

To realize the above conditions on our mechanism, we assume that the SUSY breaking field, $Z$, which has the largest vev, i.e. $|\left<F_Z\right>| \sim | m_{3/2} M_{\rm PL}|$, is charged under some symmetry so that it cannot couple to the matter fields (e.g. $S_i$, $\Psi_i$, ${\bar \Psi_i}$ and the quark and lepton chiral superfields) through operators like $K= c_{IJ} Z Q_I^\dag Q_J/M_{\rm PL} + h.c.$ and $W = c'_{IJK} Z Q_I Q_J Q_K/M_{\rm PL}$, and gauge fields through operators like $\int d^2\theta c_{k} Z W^k W^k/M_{PL} + h.c.$ ($k=\{U(1)_Y, SU(2)_L, SU(3)_c\}$).
These operators are dangerous since they will generate additional contributions to the $A_i$ with unaligned phases. If $Z$ couples to $S_i$ through these operators, it leads to a tree-level $A_i \sim m_{3/2}$ and an unconstrained phase, which clearly spoils our mechanism for solving the CP problem. If $Z$ couples to $\Psi_i$ or $\bar{\Psi}_i$, it induces $A_i$ at the one-loop level. This contribution is comparable to that from anomaly mediation in Eq.~(\ref{eq:aterm_amsb}). It, therefore, generically will spoil our mechanism. For the case where $Z$ couples to the gauge fields, $A_i$ is induced at the two-loop level which is still dangerous since it can give an $\mathcal{O}(10^{-2})$ correction to the phase of Eq.~(\ref{eq:aterm_amsb}). Lastly, when $Z$ couples to the quarks and leptons, it generates a three-loop level correction to $A_i$, which can still be dangerous. Thus, the supersymmetry breaking field must not be a true singlet.

Thus far, we have argued that operators like $K=Z |Q_I|^2 + h.c.$ must be forbidden. However, even a minimal K\"ahler potential $K=|Z|^2 + |Q_I|^2$ can be problematic. In this case, we still have effective couplings between $|Z|^2$ and $|Q_I|^2$, through the supergravity corrections to potential, $f=-3 M_{\rm PL}^2 \exp\left[-K/(3 M_{\rm PL}^2)\right]$. This induces Planck suppressed operators which generate trilinear couplings of order $A_i = -\left<Z\right>^* \left<F_Z\right>/M_{\rm PL}^2$. For $|\left<Z\right>|\sim  M_{\rm PL}$, $A_i = \mathcal{O}(m_{3/2})$ . At first glance, this might seem fine since this generates the same $A$-term for each $S_i$, phase and all. However, the anomaly mediated contribution, which always exists, is dependent on $\lambda_i^2$ and $|k_i|^2$ and so is not universal for the $A_i$. The sum of these contributions then gives a unique phase for each $A_i$, all of which can cannot be removed by field redefinitions. Thus, we restrict ourselves to models with $|\left<Z\right>| \ll 10^{-7} M_{\rm PL}$.\footnote{
If $\left<Z\right>$ is non-zero, the $F$-term of the conformal compensator $\Phi$ becomes,
\begin{equation}
\left<F\right> = \left(m_{3/2} + \frac{1}{3M_{\rm PL}^2} \left<\frac{\partial K}{\partial Z} F_Z\right>\right), \nonumber \\
\end{equation}
where $\Phi=\phi(1 + F \theta^2)$ and we choose the gauge $\phi=\exp(\left<K\right>/6 M_{\rm PL}^2)$ to recover Einstein gravity. In this case the anomaly mediated effects are changed slightly. However, this does not affect our discussion since we require the second term to be smaller than ${\mathcal O}(10^{-7}) m_{3/2}$ for any supersymmetry breaking field. 
}

For a SUSY breaking field, $Z'$, with $\left<F_Z'\right> \ll \left<F_Z\right>$,
the above arguments still apply, but the effects are suppressed by $\left<F_Z'\right>/\left<F_Z\right>$. 
This SUSY breaking field, $Z'$, is required to satisfy $|\left<Z'^* F_Z'\right>|/M_{\rm PL}^2 \ll 10^{-7} |m_{3/2}|$. 
If $Z'$ is a singlet of the symmetries,  $|\left<F_Z'\right>/M_{\rm PL}| \ll 10^{-7} |m_{3/2}|$ is further required. However this condition is very unlikely to be satisfied as this singlet field has $K= c' M_{\rm PL} Z' + h.c. \ \,  (c'=\mathcal{O}(1))$, which in turn leads to $W = c' m_{3/2} M_{\rm PL} Z'$ after a K\"ahler transformation.

If $Z'$ is a moduli-type field and the shift symmetry is not broken, it will couple to matter and SUSY breaking fields with real coefficients such as $K = c_{I}^n (Z'+Z'^*)^n |Q'_I|^2/M_{\rm PL}^{n}$, where $Q'_I$ represents the matter and SUSY breaking fields that are charged under some symmetry. Including this term in the K\"ahler potential gives
\begin{eqnarray}
-e^{K/2} \left<F_Z'\right> &=& (K^{-1})_{xx}\left< \frac{\partial K}{\partial x} W^*/M_{\rm PL}^2 \right> \nonumber\\
&+& (K^{-1})_{x i} \left< W_{i}^* + K_i^* W^*/M_{\rm PL}^2 \right>,
\end{eqnarray}
where $x=Z'+Z^{'*}$. The first term is proportional to $m_{3/2} M_{\rm PL}$ with a real coefficient, and the second term is $\sim \left<Z^* \cdot F_Z\right>/M_{\rm PL} \ll 10^{-7} m_{3/2} M_{\rm PL}$. Since the couplings to the matter fields, including $S_i$, are given by real constants this does not induce different phases for different $S_i$. 

\vspace{20pt}
Lastly, let us discuss the cosmological domain wall problem. In our model, domain walls are formed once the $Z_{3_i}$ symmetries are spontaneously broken. The domains eventually follow a scaling solution and about one domain wall will exists within a Hubble horizon~\cite{Press:1989yh,Garagounis:2002kt,Oliveira:2004he,Leite:2011sc}. In this case, their energy density can be approximated by $\sigma H^{-2}/H^{-3}=\sigma H$, where $\sigma$ is the domain wall tension and $H$ is the Hubble parameter. By comparing this to the total energy density, $3 H^2 M_{\rm PL}^2$, it turns out the domain walls dominate the energy density of the universe for
\begin{eqnarray}
H_{\rm dom} \sim \frac{\sigma}{M_{\rm PL}^2} \sim
\frac{\left<S_i\right>^3}{M_{\rm PL}^2}.
\end{eqnarray}
If the domain walls are stable enough to last until today, the energy density of the domain walls will dominate. In order to avoid this situation, we need a small $Z_{3_i}$ breaking term, which removes the degeneracy of the three vacua by an amount we label $\Delta V$. The domain walls collapse when their energy density becomes similar in size to $\Delta V$. The conditions
\begin{eqnarray}
H_{\rm col} \sim \frac{\Delta V}{\sigma} \sim \frac{\Delta V}{\left<S_i\right>^3} > H_{\rm dom},
\end{eqnarray}
should then be satisfied. Also, it is safer to require the domain walls collapse before big-bang nucleosynthesis (BBN) starts, which gives
\begin{eqnarray}
H_{\rm col} > \frac{\mathcal{O}({\rm MeV^2})}{M_{\rm PL}}.
\end{eqnarray}
$\Delta V$ should then satisfy the following condition:
\begin{eqnarray}
\Delta V > {\rm max}\left(\frac{\left<S_i\right>^6}{M_{\rm PL}^2},  \frac{\mathcal{O}({\rm MeV^2})}{M_{\rm PL}} \left<S_i\right>^3\right).
\end{eqnarray}
Let us consider the following $Z_{3_i}$ breaking term, which conserves $Z_{4R}$ symmetry,
\begin{eqnarray}
W \ni \mu^2_i S_i \to V \ni (\mu^2_i \lambda_i S_i^{2*} - m_{3/2} \mu^2_i S_i) + h.c.
\end{eqnarray}
The first term gives the dominant contribution to $\Delta V$. The conditions on $\Delta V$ lead to
\begin{eqnarray}
|\mu^2_i| > \mathcal{O}(10^{-13}){\rm GeV}^2\, \left(\frac{|\left<S_i\right>|}{10^6\, {\rm GeV}}\right)^4, \label{eq:mu2_dom}
\end{eqnarray}
and
\begin{eqnarray}
|\mu^2_i| > \mathcal{O}(10^{-22}){\rm GeV}^2\, \left(\frac{|\left<S_i\right>|}{300\, {\rm GeV}}\right).
\end{eqnarray}
The latter condition from BBN is always weaker than the former one for $\vev{S_2}$ and $\vev{S_3}$. For $\vev{S_1}$, the BBN constraint can be stronger.
On the other hand, in order not to induce dangerous CP violating phases, $|\mu^2_i|$ is bounded from above as
\begin{eqnarray}
|\mu^2_i| \ll \mathcal{O}(10^{-5}) (A_i)_{\rm AMSB} |\left<S_i\right>| \approx \mathcal{O}(10^{-3}) {\rm GeV}^2 \left(\frac{m_{3/2}}{10\,{\rm MeV}}\right) \left(\frac{|\left<S_i\right>|}{10^6\, {\rm GeV}}\right). \label{eq:mu2_cp}
\end{eqnarray}
These conditions Eqs.~(\ref{eq:mu2_dom})-(\ref{eq:mu2_cp}) are satisfied as long as $M_{\rm mess} < \mathcal{O}(10^{10}$-$ 10^{11}\,{\rm GeV})$. Here, $m_{3/2}=10\,{\rm MeV}$ is a reference value for $S_{2,3}\simeq 10^{6}$\,GeV and $m_{3/2}$ is roughly proportional to $\left<S_{2,3}\right> = M_{\rm mess}/k_{2,3}$.


We need to be a bit careful in our treatment of $S_1$ as it mixes with the Higgs doublets, which could potentially affect Higgs physics. Focusing on 
\begin{eqnarray}
	W = \frac{\lambda_1}{3} S_1^3 + k_1 S_1 H_u H_d,
\end{eqnarray}
the superpotential and soft SUSY breaking terms are exactly the same as those of $Z_3$ invariant Next-to-Minimal Supersymmetric Standard Model (NMSSM) (see Refs.~\cite{Maniatis:2009re,Ellwanger:2009dp} for a review). 
The CP-even Higgs mass matrix contains  
\begin{eqnarray}
	V &\ni& \frac{1}{2} m_h^2 h^2
	+ v k_1 \left[
	2 \mu  - \sin 2\beta (B_{\rm eff} + \lambda_1 \vev{S_1})
	\right]  h s
	+ \frac{1}{2} (4 \lambda_1^2 \vev{S_1}^2 + \frac{k_1 A'_1}{2}  \frac{v^2}{\vev{S_1}} \sin 2\beta ) s^2, \nonumber \\
	&\approx& \frac{1}{2} m_h^2 h^2
	+ (2 k_1 v \mu)\,  h s
	+ \frac{1}{2} (2 B_{\rm eff}-2A'_1)^2 s^2,
\end{eqnarray}
where $h = -\sin\beta h_u^0 + \cos\beta h_d^0$ and $m_h \approx 125\,$\,GeV; 
$\mu=k_1 \vev{S_1}, \, B_{\rm eff} = B_\mu/\mu \simeq A'_1 + \lambda_1 \vev{S_1}$, $V \ni A'_1 k_1 S H_u H_d + h.c.$; $A'_1 \sim 100\,$\GEV through gaugino loops at the SUSY mass scale; Here, we neglect $A_1$. The Higgs $B$-term, $B_{\rm eff}$, is estimated to be
\begin{eqnarray}
	B_{\rm eff} \simeq \frac{m_A^2}{\mu \tan\beta} \approx 
	296\,\GEV \times 
	\left(
	\frac{m_A}{2000\,\GEV}
	\right)^2 
	\left(\frac{300\,\GEV}{\mu}\right) 
	\left(\frac{45}{\tan\beta}\right).
\end{eqnarray}
To avoid the large mixing between $h$ and $s$, $k_1 < \mathcal{O}(0.1)$, is required, and $\lambda_1 \sim k_1$ due to $B_{\rm eff} \sim \mu$.
The mass of the singlet-like CP-odd Higgs is 
\begin{eqnarray}
	m_{a_1}^2 
&\simeq& \frac{k_1 \sin 2 \beta}{2}(4 B_{\rm eff} - 3 A'_1) \frac{v^2}{\vev{S_1}}-3 A_{1} \lambda_1 \vev{S_1}	\\
	&\approx& -3 A_{1} \lambda_1 \vev{S_1} \approx 3 (B_{\rm eff}-A'_1) |A_{1}|,
\end{eqnarray}
which can be larger than $\mathcal{O}(1)$\,GeV. This $a_1$ as well as the imaginary parts of $S_{2,3}$ (i.e. axion-like particles) can safely decay before the BBN.

\section{Discussion and conclusions}

We presented a CP-safe version of the gauge mediated model in \cite{Bhattacharyya:2018inr} by extending the mechanism proposed in Ref.~\cite{Evans:2010ru} to include anomaly mediation effects. First, we showed that large portions of the parameter space are still consistent with the $g_\mu-2$ and LHC constraints. For heavy sleptons, the CP-odd Higgs and the Heavy CP-even Higgs tend to be light and, therefore, discoverable at the LHC. This analysis is for a generic minimal gauge mediation model with messengers in the fundamental representation of $SU(5)$.  

We also showed that the dangerous CP violating phases can be drastically suppressed by applying the mechanism first proposed in \cite{Evans:2010ru}. This mechanism allows us to dynamically suppress the phases of all relevant supersymmetry breaking parameters. This mechanism aligns the soft mass phases with that of $m_{3/2}$ through anomaly mediated $A$-terms in the messengers and Higgs bosons potential. Then, when the gravitino phase is rotated away all soft masses become real to a high level of precision leaving no dangerous CP violating phase. Furthermore, in Appendix~\ref{sec:a1}, we give a consistent ultraviolet model for this mechanism.

\section*{Acknowledgments}
We thank Andre Lessa for useful discussions about ATLAS-SUSY-2018-41 constraints and the usage of {\tt SmodelS}.
N. Y. is supported by a start-up grant from Zhejiang University. J. L. E. is supported by a start-up grant from Shanghai Jiao-Tong University.
T. T. Y. is supported in part by the China Grant for Tal ent Scientific Start-Up Project and by Natural Science Foundation of China (NSFC) under grant No. 12175134 as well as by World Premier International Research Center Initiative (WPI Initiative), MEXT, Japan.

\appendix
\section{Generation of soft SUSY breaking masses beside the gauge mediation}\label{sec:a1}

Let us first discuss the tachyonic mass generation for $S_i~ (i=1 \dots 3)$, which determines the messenger mass scale, the messenger $B$-terms (therefore $\Lambda_{L,D})$, the $\mu$-term and the Higgs $B$-term. 
For this purpose, we introduce new Yukawa couplings for the $S_i$,
\begin{equation}
    W=\frac{k'_i}{2}S_i E_i^2,
\end{equation}
where $E_i$ are gauge singlet fields. The tachyonic masses for $S_i$ are generated by one-loop diagrams involving $E_i$, if the soft SUSY breaking mass squared of $E_i$ is positive. This is analogous to how the stop loops generate a tacyonic Higgs soft SUSY breaking mass which generates electroweak symmetry breaking radiatively. 

A positive supersymmetry breaking mass is generated for $E_i$ from the following superpotential:
\begin{eqnarray}
W = \frac{\kappa}{2} Z\Psi^2 + M_0\Psi{\bar\Psi} + M_i E'_i\bar{E}'_i + g_i E_i {E}_i'\Psi + W_{\rm IYIT}, \label{eq:the_model}
\end{eqnarray}
where $M_0 \gg M_i$ is assumed, and $\Psi$, $\bar\Psi$, $E'_i$ and $\bar{E}'_i$ are all gauge singlet fields. A consistent set of charge assignment for these fields under $Z(3)_1\times Z(3)_2\times Z(3)_3 \times Z_{4R}$ is shown in table \ref{tb:charges}. 
We assume the non-anomalous $Z_{4R}$ is a discrete gauge symmetry while the others are global symmetries.
The charges of the matter multiplets in the standard model sector under $Z(3)_1\times Z(3)_2\times Z(3)_3\times Z_{4R}$ are ${\bf 5}^*(-2,0,0,1), {\bf 10}(-1,0,0,1)$ and $N(0,0,0,1)$. Here, we again use $SU(5)$ notation for simplicity and $N$ is for the right-handed  neutrinos. These charge assignments are consistent with the standard model Yukawa interactions.  
\begin{table*}[t]
\caption{Charge assignment for all the fields. \label{tb:charassign} }
\begin{center}
\begin{tabular}{|c|c|c|c|c|c|}
\hline 
Operators & $Z_{3_1}$ & $Z_{3_2}$ & $Z_{3_3}$ & $Z_{4R}$ & $Z_4$\\  

\hline  $Z$ & 0 & 0 & 0 & 2 & 2\\
\hline  $Q_a$ & 0 & 0 & 0 & 0 & 1\\
\hline $\Psi$ & 0 & 0  & 0& 0 & -1\\
\hline ${\bar \Psi}$ &0&0&0&2 & 1 \\
\hline $E'_1$ & 2 & 0 & 0 & 2 & 1\\
\hline ${\bar E}'_1$ & 1 & 0 & 0 & 0 & -1 \\
\hline $E'_2$ & 0 & 2 & 0 & 2 & 1\\
\hline ${\bar E}'_2$ & 0 & 1 & 0 & 0 & -1\\
\hline $E'_3$ & 0 & 0 & 2 & 2 & 1\\
\hline ${\bar E}'_3$ & 0 & 0 & 1 & 0 & -1\\
\hline $E_1$ & 1 & 0 & 0 & 0 & 0\\
\hline $E_2$ & 0 & 1 & 0 & 0 & 0\\
\hline $E_3$ & 0 & 0 & 1 & 0 & 0\\
\hline $H'_u$ & 2 & 0 & 0 & 0 & 0 \\
\hline ${\bar H}'_u$ &1 &0&0& 2 & 0\\
\hline $H'_d$ & 0 & 0 & 0 & 0 & 0\\
\hline ${\bar H}'_d$ &0 & 0& 0& 2 &0\\
\hline  $S_1$ & 1 & 0 &0 &2 & 0\\
\hline  $S_2$ & 0 & 1 &0 &2 & 0\\
\hline  $S_3$ & 0 & 0 &1 &2 & 0 \\
\hline $\Psi_{\bar D} \Psi_D$ & 0 & 0 & 2 & 0 &0\\
\hline  $\Psi_{L} \Psi_{\bar L}$ & 0 & 2 &0 &0 &0\\
\hline  $H_u$ & 2 & 0 & 0 & 0 & 0\\
\hline  $H_d$ & 0 & 0 & 0 & 0  &0\\
\hline
\end{tabular}
\end{center}
\label{tb:charges}
\end{table*}%

The last term in Eq.(\ref{eq:the_model}), $W_{\rm IYIT}$, represents the effective superpotential for dynamical SUSY breaking based on an IYIT model~\cite{Izawa:1996pk,Intriligator:1996pu} with an $SP(1)=SU(2)$ gauge theory. After the quark superfields condense, the superpotential takes the following form:
\begin{equation}
    W_{\rm IYIT}=\frac{1}{2} (Q_1Q_2 +Q_3Q_4)Z =\Lambda^2 Z.
\end{equation}
Here, $Q_a ~ (a=1 \dots 4)$ are doublet quarks, and $\Lambda$ is the dynamical scale. This superpotential breaks supersymmetry, i.e. $F_Z=\Lambda^2$.
The IYIT model with the $SP(1)$ gauge symmetry has a discrete $Z(4)$ symmetry, where the doublet quarks $Q_a$ carry $+1$ charges and $Z$ has $+2$. Accordingly, the gauge singlet fields, $\Psi$, ${\bar \Psi}$, $E'_i$, ${\bar E}'_i$ carry  charges $-1$, $+1$, $+1$ and $-1$, respectively. All other fields carry a charge of zero under this $Z(4)$ symmetry. Importantly, this discrete $Z(4)$ symmetry prohibits a linear term involving ${\bar \Psi}$ and mass mixing between $E'_i$ and $E_i$.
The $E_i$ and $\bar{E}'_i$ can still decay before BBN through Planck-suppressed operators which break $S(3)_i$ and $Z(4)$ through operators such as $K=E_i^\dagger H_u H_d/M_{\rm PL}$, $\bar{E}^{'\dag}_i H_u H_d/M_{\rm PL}$.

The positive squared masses for $E'_i$, which are generated after integrating out $\Psi, {\bar \Psi}$, are estimated using the one-loop formula for the effective K\"ahler potential~\cite{Grisaru:1996ve}. The field dependent mass matrix for $\Psi, {\bar \Psi}$ and $E'_i$ is 
\begin{eqnarray}
\mathcal{M} &=&
\left(
\begin{array}{ccc}
\kappa Z & M_0 & g_i E_i \\
M_0 & 0 & 0  \\
g_i E_i & 0 &0 \\
\end{array}
\right).
\end{eqnarray}
The matrix, $\mathcal{M}^\dag \mathcal{M}$, has two non-zero eigenvalues, $m_+^2$ and $m_-^2$, which are 
\begin{eqnarray}
m_{\pm}^2 &=& \frac{1}{2} \left(2 |M_0|^2+ 2 |g_i E_i|^2  + |\kappa Z|^2 \pm \sqrt{|\kappa Z|^2}\sqrt{|\kappa Z|^2  +4(|M_0|^2 +|g_i E_i|^2)}
\right).\nonumber \\ 
\end{eqnarray}
Using this mass matrix, the effective K\"ahler potential can be written as
\begin{eqnarray}
K_{\rm eff} &=& - \frac{1}{32\pi^2} {\rm Tr} 
\left(
\mathcal{M}^\dag \mathcal{M} \ln \frac{\mathcal{M}^\dag \mathcal{M}}{e^{c_r}Q^2}
\right) \nonumber \\
&=& - \frac{1}{32\pi^2}  
\left(
m_+^2 \ln \frac{m_+^2}{e^{c_r}Q^2}
+ m_-^2 \ln \frac{m_-^2}{e^{c_r}Q^2}
\right).
\end{eqnarray}
where $Q$ is the renormalization scale and $c_r$ is a scheme dependent constant. We then obtain the following soft SUSY breaking masses for $m_{E_i}^2$:
\begin{eqnarray}
m_{E_i}^2 &\simeq& -\left. \frac{\partial^2 K_{\rm eff}}{\partial (|E_i|^2) \partial (|Z|^2)} \right|_{|E_i|^2=|Z|^2=0}|F_Z|^2 = \frac{|g_i \kappa|^2}{32\pi^2} \left|\frac{F_Z}{M_0}\right|^2.
\end{eqnarray}
Thanks to the fact that this mass squared is positive, a tachyonic mass for the $S_i$ is generated by $E_i$ loops:\footnote{
The operator in the K\"ahler potential giving the tachyonic mass, $K\sim 1/(16\pi^2)^2 |Z|^2|S_i|^2/M_0^2$, also induce $A$-terms for $S_i$, with $\left<Z\right>=\mathcal{O}(m_{3/2})$ induced by effects from supergravity. However, $\left<Z F_Z^*\right>$ is real, and it does not changes the phase direction of $\left<S_i\right>$.
}
\begin{equation}
    m^2_{S_i}\simeq -\frac{|k'_i|^2}{16\pi^2}m_{E_i}^2 \ln \frac{|M_0|^2}{|m_{E_i}^2|}.
\end{equation}

To generate the soft masses for $H_u$ and $H_d$, we introduce additional pairs of massive Higgs multiplets $H'_u$ and ${\bar H'_u}$, and $H'_d$ and ${\bar H'_d}$. The $Z(4)$ charges of
$H'_u$, ${\bar H'_u}$, $H'_d$ and ${\bar H'_d}$ are $-1$, $+1$, $-1$ and $+1$, respectively.
The superpotential for these massive Higgs multiplets is given by 
\begin{equation}
    W=\lambda_u H_u {\bar H'_u}\Psi + \lambda_d H_d {\bar H'_d}\Psi + M_u {\bar H'_u}H'_u + M_d{\bar H'_d}H'_d.
\end{equation}
The superpotential which generates the Higgs soft masses is then the same as Eq. (\ref{eq:the_model}) with $E_i \to H_{u,d}$, $E'_i \to \bar{H}'_{u,d}$, and $\bar{E}'_i \to {H}'_{u,d}$. In the limit $M_u \sim M_d \ll M_0$, we then get
\begin{eqnarray}
\Delta m_{H_{u,d}}^2 &\simeq& \frac{|\lambda_{u,d} \kappa|^2}{32\pi^2} \left|\frac{F_Z}{M_0}\right|^2.
\end{eqnarray}

\section{Some motivated extensions} \label{ap:ext}
Let us consider the following superpotential:
\begin{eqnarray}
W = \lambda_{X_0} X_0 (X \bar X -  v^2) + \lambda_X X \psi \bar{\psi},
\end{eqnarray}
where $X_0$ and $\Psi \bar{\Psi}$ has a $U(1)_R$ (or a discrete R) charge of 2. We consider the case where $X$ and $\bar X$ break some symmetries like a global $U(1)_{\rm PQ}$ or gauged $U(1)_{B-L}$. As we will show, 
\begin{eqnarray}
B_{\psi \bar{\psi}}/M_{\psi \bar{\psi}} = {\rm real\  constant} \times m_{3/2} = {\rm real},
\end{eqnarray}
where the $B$-term and SUSY preserving mass terms are given by
\begin{eqnarray}
\mathcal{L} = B_{\psi \bar{\psi}}\psi \bar{\psi} + h.c., \ \  M_{\psi \bar{\psi}} =  \lambda_X \left<X\right>.
\end{eqnarray}
Because the ratio of these terms is real, this extension has no effect on our mechanism for suppressing CP phases. If this were not the case, these extensions could generate dangerously large CP phases. For instance, in the case of a $U(1)_{\rm PQ}$, a gluino mass is induced at the one-loop level and in turn generate $S_3^3$ potential term at the three-loop level. For $U(1)_{\rm B-L}$ breaking, an $A$-term is induced for $L H_u {\bar N}$ at the one-loop level and $S_1^3$ is induced at the three-loop level as well. 

The B-term is given by
\begin{eqnarray}
\mathcal{L} = -\left[(\lambda_{X_0} X_0 \bar{X})^* + a' m_{3/2} X\right] \lambda_X \psi \bar{\psi} + h.c.,
\end{eqnarray}
where $a'$ is a real coefficient given by anomaly mediation. First, we assume $\left<X\right> \sim \left<\bar X\right> \sim v$, which is satisfied when the soft SUSY breaking masses for $X$ and $\bar {X}$ are the same order, $\mathcal{O}(m_{3/2})$, and from the minimization conditions we get ${\rm Arg}\left<X \bar{X} \right> = {\rm Arg}(v^2)$.
The potential for $X_0$ is 
\begin{eqnarray}
V = m_{X_0}^2 |X_0^2| + (- 2m_{3/2} \lambda_{X_0} v^2 X_0 + a''  m_{3/2} \lambda_{X_0} X_0 X \bar{X}) + h.c.,
\end{eqnarray}
where $a''$ is a real coefficient induced by anomaly mediation and $m_{X_0}^2=2|\lambda_{X_0}|^2(|X|^2+|\bar X|^2)$ with $|X|=|\bar X|=v$. The VEV of $X_0$ is fixed to be 
\begin{eqnarray}
\left<|X_0|\right> \simeq | \lambda_{X_0} m_{3/2}| \ \ {\rm and} \ {\rm Arg} ( \left<X_0\right>) = - {\rm Arg}(v^2) - {\rm Arg}(\lambda_{X_0})~,
\end{eqnarray}
so that 
\begin{eqnarray}
{\rm Arg}\left((\lambda_{X_0} X_0 \bar{X})^*\right)= {\rm Arg}(\left<X\right>),
\end{eqnarray}
and $B_{\psi \bar{\psi}}/M_{\psi \bar{\psi}}$ is real.

\section{The tree-level contribution to $A_1$}\label{ap:treeB}
Here, we show a model which generates a tree-level contribution to $A_1$ with the opposite sign of $(A_1)_{\rm AMSB}$. The superpotential is given by
\begin{eqnarray}
W = k_{Y} S_1^2 Y/2 + M_Y Y \bar Y + \lambda_1 S_1^3/3.
\end{eqnarray}
The scalar potential is 
\begin{eqnarray}
V &=& |M_Y|^2 |Y|^2 + |S_1|^2 |\lambda_1 S_1 + k_Y Y|^2 + \frac{1}{4}|k_Y S_1^2 + 2 M_Y \bar{Y}|^2 \nonumber \\
&-& (m_{3/2} M Y \bar Y + h.c.) + \mathcal{O}(M_{\rm PL}^{-2}).
\end{eqnarray}
By integrating out $Y$ and $\bar{Y}$ using the equations of motion, $\frac{\partial V}{\partial Y}=0$ and $\frac{\partial V}{\partial \bar{Y}}=0$, the effective scalar potential is
\begin{eqnarray}
V_{\rm eff} &=& \lambda_1^2 |S_1|^4 \frac{|M_Y|^2}{|M_Y|^2 + |k_Y|^2 |S_1|^2} - 
\left[
\frac{3}{2} m_{3/2} \lambda_1 S_1^3/3 \frac{|k_Y|^2 |S_1|^2}{|M_Y|^2 + |k_Y|^2 |S_1|^2} + h.c.
\right] \nonumber \\
&=& \lambda_1^2 |S_1|^4 \frac{|M_Y|^2}{|M_Y|^2 + |k_Y|^2 |S_1|^2} - 
\left[
(A_1)_{\rm tree} \lambda_1 S_1^3/3 + h.c. 
\right], \label{eq:a1tree}   
\end{eqnarray}
where we have neglected $\mathcal{O}(m_{3/2}^2)$ terms. It can be seen that the sign of $S_1^3$ is opposite to AMSB. The $A$-term in Eq.~(\ref{eq:a1tree}) can be larger than $(A_1)_{\rm amsb}$. For instance, by taking $|M_Y|^2 = 10 |k_Y|^2 |S_1|^2$, the $A$-term is $\approx -0.14 m_{3/2}$, which is larger than that of AMSB.

\providecommand{\href}[2]{#2}\begingroup\raggedright\endgroup

\end{document}